\documentclass[aps,prb,twocolumn,groupedaddress,showpacs]{revtex4}
\usepackage{graphicx}
\usepackage{amsmath}
\begin{document}

\title{Two-channel Kondo phases and frustration-induced transitions in triple quantum dots}

\author{Andrew K. Mitchell and David E. Logan}

\affiliation{Department of Chemistry, Physical and Theoretical Chemistry, Oxford University, South Parks Road, Oxford OX1 3QZ, United Kingdom}

\date{September 19, 2009}

\begin{abstract}
We study theoretically a ring of three quantum dots mutually coupled by antiferromagnetic exchange interactions, and tunnel-coupled to two metallic leads: the simplest model in which the consequences of local frustration arising from internal degrees of freedom may be studied within a 2-channel environment. Two-channel Kondo (2CK) physics is found to predominate at low-energies in the mirror-symmetric models considered, with a residual spin-$\tfrac{1}{2}$ 
overscreened by coupling to both leads. It is however shown that two distinct 2CK phases, with different ground state parities, arise on tuning the interdot exchange couplings. In consequence a frustration-induced quantum phase transition occurs, the 2CK phases being separated by a quantum critical point for which an effective low-energy model is derived. Precisely at the transition, parity mixing of the quasi-degenerate local trimer states acts to destabilise the 2CK fixed points; and the critical fixed point is shown to consist of a free pseudospin together with effective 1-channel spin quenching, itself reflecting underlying channel-anisotropy in the inherently 2-channel system. Numerical renormalization group techniques and physical arguments are used to obtain a detailed understanding of the problem, including study of both thermodynamic and dynamical properties of the system.

\end{abstract}


\maketitle
\section{INTRODUCTION}\label{intro}
Quantum dot devices have recently been the focus of intense investigation, due to the impressive experimental control available in manipulating the microscopic interactions responsible for Kondo
physics\cite{hewson,rev:mesoscopic,rev:Pustilnik,1dot:Goldhaber,1dot:Cronenwett}. In particular
the classic spin-$\tfrac{1}{2}$ Kondo effect\cite{hewson} --- in which a localized spin is fully screened by coupling to itinerant conduction electrons in a single attached metallic lead --- has 
now been widely studied experimentally (see \emph{eg} [\onlinecite{1dot:Goldhaber,1dot:Cronenwett,artmol,A-B:LPK,nanotube:Nygard}]). 

But arguably the most diverse and subtle Kondo physics results from the \emph{interplay} between internal spin and orbital degrees of freedom in \emph{coupled} quantum dots. A wide range of strongly correlated electron behavior is accessible in such systems, with variants of the standard Kondo effect\cite{rev:nutshell,bordaDQD,ferrodots,galpinccdqd,3dot1ch:us,Martins2009,OguriTQD2009}, quantum phase transitions\cite{galpinccdqd,3dot1ch:us,ingersentDQD,2ck:propose1,zitkoFL_NFL} and non-Fermi liquid phases\cite{2imp:prl,2ck:propose2,2ck:DDcrit,2ck:NFLinQD} having been studied theoretically. 
In particular, both double\cite{kikoinDQD,vojta2spin,bordaDQD,galpinccdqd,AKMccDQD,andersgalpinnano,lopez,zarandDQD,ingersentDQD} and triple\cite{lazarovitsadatom,hewsonTQD,zitkoTQD,ferrodots,lobosTQD,wangtqd,delgadoTQD,zitkoTQD2ch,TQD:ingersent,Martins2009,OguriTQD2009} quantum dots have been shown to demonstrate low-temperature behavior quite different from their single-dot counterparts\cite{rev:nutshell}. 
Advances in nanofabrication techniques\cite{1dot:Goldhaber,1dot:Cronenwett,artmol,A-B:LPK,nanotube:Nygard,blick2dot,QPT:science,gaudreau3dot,ludwig3dot,vidan3dot,rogge3dot,nanotube:Grove,2ckexpt:potok}, and atomic-scale manipulation using scanning tunneling microscopy\cite{Crtrimer:crommie,Cotrimer:uchihashi}, now allow for the construction of coupled quantum dot
structures, in which the geometry and capacitance of the dots can be controlled, and their couplings fine-tuned\cite{rev:mesoscopic,gaudreau3dot}. Experimental access to a rich range of Kondo and related physics is thus within reach.

One of the most delicate effects, however, arises in the two-channel Kondo (2CK) model proposed by Nozi\`{e}res and Blandin\cite{2CK:nozieres}, which describes a spin-$\tfrac{1}{2}$ symmetrically coupled to two independent metallic leads. The standard, strong coupling Fermi liquid fixed point common in single-channel models is destabilized here. Much studied theoretically (for a review see [\onlinecite{2ck:cox_zaw}]), the dot spin in the 2CK model is overscreened at low temperatures, with each channel competing to compensate the local moment. The nontrivial intermediate-coupling fixed point which controls the low-temperature behavior of the system has a number of unusual non-Fermi-liquid (NFL) properties\cite{TD:Destri,TD:Tsvelick,TD:Sacramento,2ck:cox_zaw}, including a residual entropy of $\tfrac{1}{2}\ln (2)$ and a magnetic susceptibility which diverges logarithmically at low temperature.

The key ingredient of this 2CK physics is the suppression of charge transfer between the two symmetrically coupled leads\cite{2CK:nozieres,2ck:cox_zaw,2ck:affleck2} --- the necessity in essence of a central \emph{spin}.  Experimental realizations of 2CK physics have been variously sought in \emph{eg} heavy fermion systems containing Uranium\cite{U:cox,U:seaman,U:andrei},  scattering from 2-level systems using ballistic metal point contacts\cite{pointcontact:Ralph1,pointcontact:Ralph2}, systems with tunneling between nonmagnetic impurities in metals\cite{nonmag:Vladar1,nonmag:Vladar2}, and from impurities in graphene\cite{graph:baskaran}; albeit that the interpretation of observed behavior in terms of 2CK physics is not always unambiguous.

Recently however, Potok \emph{et al} have demonstrated\cite{2ckexpt:potok} clear two-channel behavior in a coupled quantum dot device in which one small and one large dot are tunnel-coupled, with the small dot also coupled to a metallic lead. The large dot acts as an interacting second lead, but is fine-tuned to the Coulomb blockade regime so that charge transfer is energetically disfavoured. A small degree of inter-lead charge transfer must nonetheless 
occur, so ultimately the system is a Fermi liquid, with the 2CK fixed point destabilized below some low-temperature scale (which crossover has been widely studied theoretically\cite{2ck:affleck2,zarandDQD,zitkoFL_NFL}, as has the instability of the 2CK fixed point to channel
asymmetry\cite{2CK:nozieres,2ck:affleck1,AndreiJerez2CK,TD:Sacramento}). But at finite temperatures, 2CK physics is clearly observed\cite{2ckexpt:potok}.

Studying coupled quantum dot systems in a two-channel environment has attracted considerable theoretical interest recently\cite{2imp:ingersent,2ck:propose1,2ck:DDcrit,2ck:mag,2ck:Kuz,zitkoFL_NFL,zitkoTQD2ch}, in part because NFL states are accessible. The two-impurity two-channel Kondo model\cite{2imp:jayaprakash,2imp:jones,2imp:ingersent} (where two antiferromagnetically coupled dots are each coupled to their own lead) is a prime example, in which the tendency to form a local singlet state on the two dots competes with the formation of two separated single-channel Kondo states\cite{2imp:jayaprakash,2imp:jones,2imp:ingersent}. The quantum critical point separating these phases is again the 2CK fixed point\cite{2imp:affleck,2imp:ganFP}. Triple quantum dot (TQD) models with three dots coupled to two leads in a mirror-symmetric fashion have also been studied\cite{TQD:Moustakas,2ck:Kuz,zitkoFL_NFL,zitkoTQD2ch,Martins2009,OguriTQD2009}, 
recent work\cite{zitkoTQD2ch} by \v{Z}itko and Bon\v{c}a showing in particular that a range of 
fixed points familiar from simpler quantum impurity models are accessible in a ring model. Indeed both two-channel and two-impurity Kondo effects are realised, on tuning the interdot couplings as a third dot is coupled to the 
two-impurity two-channel system\cite{zitkoTQD2ch}.

\begin{figure}[t]
\includegraphics[height=3.4cm]{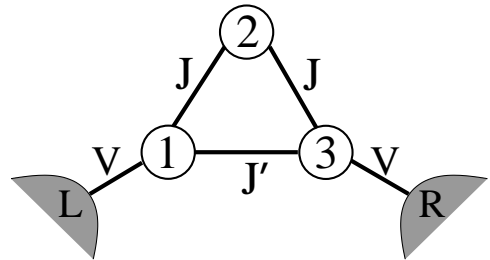}
\caption{\label{dots}
Schematic of the two-channel quantum dot trimer.
}\end{figure}

TQDs arranged in a ring geometry also provide\cite{vidan3dot,3dot1ch:us,zitkoTQD2ch} the simplest and most direct means of studying the effect of local frustration on Kondo physics. In mirror-symmetric systems, all dot states can be classified by their parity under left$\leftrightarrow$right interchange\cite{ferrodots,3dot1ch:us}. This symmetry permits a level crossing of states in the isolated trimer on varying the interdot couplings, with a pair of degenerate doublets comprising the ground state when all dots are equivalent. It was recently shown\cite{3dot1ch:us} 
that this level-crossing is preserved in the full many-body system when a single lead is attached. The situation is however more subtle on coupling the trimer to two leads, which is the focus of the present paper. We study a two-channel TQD ring model, shown schematically in Fig.~\ref{dots} (and discussed below), as a function of the interdot exchange couplings; using the density matrix extension\cite{asbasis,fdmnrg} of Wilson's numerical renormalization group (NRG) technique\cite{nrgreview,KWW,nrg_rev}.

We show that 2CK physics predominates in this model, but that two \emph{distinct} 2CK phases must in fact arise since local trimer states of different parity can couple to the leads. In consequence, one expects a quantum phase transition to occur between the two 2CK phases.  This is indeed shown to arise, with the 2CK phases separated by a nontrivial quantum critical point, the nature of which is uncovered explicitly and analysed in detail.

The paper is organized as follows. In Sec.~\ref{model} we discuss the two-channel TQD Hamiltonian, and develop low-energy effective models to describe the behavior of the system when deep in each 2CK phase. Symmetry arguments 
indicate the presence of a quantum phase transition near to the point of inherent magnetic frustration in the TQD, and an effective low-energy model valid in the vicinity of the transition is derived. Sec.~\ref{2CK} presents 
NRG results for the full system, considering both thermodynamics and dynamical properties of the 2CK phases. The transition itself is investigated in Sec.~\ref{QPT}, and the nature of the critical point elucidated, employing heuristic physical arguments in addition to direct calculation. In Sec.~\ref{red} the effective low-energy model describing the transition is itself studied directly using NRG. The paper concludes with a brief summary.

Before proceeding, we point out that the stability of 2CK physics in the model studied here is of course delicate, just as it is for the standard two-channel Kondo  model\cite{2CK:nozieres,2ck:cox_zaw,2ck:affleck2}. A small degree of charge transfer, which would arise in a real TQD device from interdot tunnel-couplings (as opposed to exchange couplings), will ultimately lead to a crossover\cite{2ck:affleck2,zarandDQD,zitkoFL_NFL} from a 2CK to a stable Fermi liquid fixed point, below some characteristic low-temperature scale. The same situation is of course well known to occur for channel anisotropy\cite{2CK:nozieres,2ck:affleck1,AndreiJerez2CK,TD:Sacramento}, as would
arise in the TQD model upon destruction of overall left$\leftrightarrow$right symmetry via \emph{eg} different exchange couplings $J_{23} \neq J_{21}$ (see Fig.~\ref{dots}).


\section{2CK trimer model}\label{model}
We consider a system of three (single-level) quantum dots, arranged in a triangular geometry, as illustrated in Fig.~\ref{dots}. Dot `2' is coupled to dots `1' and `3', which are coupled to each other and to their own metallic lead. We model the central dot 2 strictly as a \emph{spin}-$\tfrac{1}{2}$ to prevent inter-lead charge transfer, but dots 1 and 3 are Anderson-like sites, permitting variable occupation. Tunneling is allowed between these terminal dots and their connected leads, but the dots are coupled to each other by an antiferromagnetic (AF) exchange interaction to form a Heisenberg ring. We focus explicitly on a system tuned to left/right mirror symmetry (see Fig.~\ref{dots}), with Hamiltonian $H=H_0+H_{\text{hyb}}+H_{\text{tri}}$. Here $H_0=\sum_{\alpha,\textbf{k},\sigma}\epsilon^{\phantom{\dagger}}_{\textbf{k}} a^{\dagger}_{\alpha \textbf{k} \sigma}a^{\phantom{\dagger}}_{\alpha \textbf{k} \sigma}$ refers to the two equivalent non-interacting leads ($\alpha=L,R$), which are tunnel-coupled to dots 1 and 3 via $H_{\text{hyb}} = \sum_{\textbf{k},\sigma} V(a^{\dagger}_{L\textbf{k} \sigma}c^{\phantom{\dagger}}_{1 \sigma}+a^{\dagger}_{R\textbf{k} \sigma}c^{\phantom{\dagger}}_{3 \sigma}+\text{H.c.})$. $H_{\text{tri}}$ describes the trimer itself, with exchange couplings $J$ and $J'$,
\begin{equation}
\label{Hfull}
\begin{split}
H_{\text{tri}} = &J \hat{\textbf{S}}_2 \cdot (\hat{\textbf{S}}_1 + \hat{\textbf{S}}_3) + J'\hat{\textbf{S}}_1 \cdot \hat{\textbf{S}}_3\\
& +\epsilon(\hat{n}_1+\hat{n}_3)+U(\hat{n}_{1\uparrow}\hat{n}_{1\downarrow}+\hat{n}_{3\uparrow}\hat{n}_{3\downarrow}) 
\end{split}
\end{equation}
where $\hat{\textbf{S}}_i$ is a spin-$\tfrac{1}{2}$ operator for dot $i$. For dots $i=1,3$, $\hat{n}_i=\sum_{\sigma}\hat{n}^{\phantom{\dagger}}_{i\sigma}=\sum_{\sigma}c^{\dagger}_{i \sigma}c^{\phantom{\dagger}}_{i \sigma}$ is the number operator, $\epsilon$ is the level energy and $U$ is the intradot Coulomb repulsion. The full Hamiltonian $H$ is thus invariant under simultaneous $1\leftrightarrow 3$ and $L\leftrightarrow R$ permutation.


\subsection{The isolated trimer}\label{TQD}
We are interested in the TQD deep in the $\mathcal{N}=3$ Coulomb blockade valley, where each dot is in practice singly occupied. To this end we consider explicitly $\epsilon=-\tfrac{1}{2}U$, the $\mathcal{N}=2$ or $4$ states being much higher ($\sim\tfrac{1}{2}U$) in energy. $H_{\text{tri}}$ (and the full $H$) is then particle-hole symmetric, although this is incidental: we require only that the singly-occupied manifold of TQD states lies lowest. This manifold comprises two doublets and a spin quartet, which is always $\max (J,J')$ higher than the doublets for AF exchange couplings.

For any $J$, $J'$, the lowest doublets of the isolated TQD are
\begin{equation}
\label{states}
\begin{split}
|+;S^z \rangle &= \hat{S}_{2}^{\sigma} \tfrac{1}{\sqrt{2}} \left( c_{1 \uparrow}^{\dagger} c_{3 \downarrow}^{\dagger} + c_{3 \uparrow}^{\dagger} c_{1 \downarrow}^{\dagger} \right) |\text{vac} \rangle \\
|-;S^z \rangle &= \tfrac{\sigma}{\sqrt{6}} \left[ \hat{S}_{2}^{\sigma} (c_{1 \uparrow}^{\dagger} c_{3 \downarrow}^{\dagger} - c_{3 \uparrow}^{\dagger} c_{1 \downarrow}^{\dagger} ) - 2\hat{S}_{2}^{-\sigma} c_{1 \sigma}^{\dagger} c_{3 \sigma}^{\dagger} \right] |\text{vac} \rangle
\end{split}
\end{equation}
with $S^z=\tfrac{\sigma}{2}$ and $\sigma = \pm$ for spins $\uparrow$$/$$\downarrow$, and $\hat{S}_{2}^{\sigma}\equiv \hat{S}_{2}^{\pm}$ the raising/lowering operator for the spin on dot 2. $|\text{vac}\rangle=\sum_{\sigma_2}|-;\sigma_2;-\rangle$ defines the `vacuum' state of the local (dot) Hilbert space, in which dots 1 and 3 are unoccupied, and dot 2 carries a spin-$\tfrac{1}{2}$.

The energy separation of the two doublets is $E_{\Delta}=E_+-E_-=J-J'$, with $|+;S^z\rangle$ the ground state of the isolated TQD for $J'>J$ and $|-;S^z\rangle$ lowest for $J'<J$. When $J'$ is dominant, dots 1 and 3 naturally lock up into a singlet (see $|+;S^z \rangle$, Eq.~\ref{states}), leaving a free spin on dot 2; with $\langle\hat{\textbf{S}}_1\cdot\hat{\textbf{S}}_3\rangle=-\tfrac{3}{4}$ and $\langle\hat{\textbf{S}}_1\cdot\hat{\textbf{S}}_2\rangle=\langle\hat{\textbf{S}}_3\cdot\hat{\textbf{S}}_2\rangle=0$. For $J'<J$ by contrast, dots 1 and 3 are now in a \emph{triplet} configuration ($|-;S^z \rangle$ in Eq.~\ref{states}), with $\langle\hat{\textbf{S}}_1\cdot\hat{\textbf{S}}_3\rangle=+\tfrac{1}{4}$ and $\langle\hat{\textbf{S}}_1\cdot\hat{\textbf{S}}_2\rangle=\langle\hat{\textbf{S}}_3\cdot\hat{\textbf{S}}_2\rangle=-\tfrac{1}{2}$.

The states are degenerate precisely at $J'=J$, reflecting the inherent magnetic frustration at that point. A level crossing of the doublets at $J'=J$ is permitted because each has different symmetry under $1\leftrightarrow 3$ permutation. We define a parity operator $\hat{P}_{i,j}$ which exchanges orbital labels $i$ and $j$ (as discussed further in the Appendix). From Eq.~\ref{Hfull} it is clear that $\hat{P}_{1,3}$ commutes with the isolated TQD Hamiltonian, $[\hat{P}_{1,3},H_{\text{tri}}]=0$. Thus all states of $H_{\text{tri}}$ can be classified according to $1\leftrightarrow 3$ parity, the eigenvalues of $\hat{P}_{1,3}$ being $\pm 1$ only (since $\hat{P}_{1,3}^{2}=1$). 
In the spin-only (`singly-occupied') regime, the parity operator may be expressed concisely as \cite{dirac} $\hat{P}_{1,3}=(1-\hat{\textbf{S}}^2_{13})$, with $\hat{\textbf{S}}_{13}=(\hat{\textbf{S}}_{1}+\hat{\textbf{S}}_{3})$ the total spin of dots 1 and 3. Thus $\hat{P}_{1,3}|\pm;S^z \rangle=\pm|\pm;S^z \rangle$ describes the parity of the doublet states of $H_{\text{tri}}$. The full lead-coupled Hamiltonian $H$ is not of course invariant to $1\leftrightarrow 3$ interchange alone, but rather to simultaneous exchange of the $1,3$ dot labels \emph{and} the left/right leads (embodied in
$\hat{P}_{L,R}=\hat{P}_{1,3}\prod_{\textbf{k}}\hat{P}_{L\textbf{k},R\textbf{k}}$, see Appendix); which we
refer to as `overall $L\leftrightarrow R$' symmetry.


\subsection{Effective low-energy models}
\label{eff models}
On tunnel-coupling to the leads, effective models describing the system on low-energy/temperature scales in the
${\cal{N}}=3$-electron valley of interest may be obtained by standard Schrieffer-Wolff
transformations\cite{hewson,SW}, perturbatively eliminating virtual excitations into the ${\cal{N}}=2$- and
$4$-electron sectors of $H$ to second order in $H_{\text{hyb}}$ (and neglecting retardation effects~\cite{hewson} as usual).
The calculations are lengthy, so rather than giving full details we sketch below a somewhat simplified, but physically more transparent, account of the key results (Eqs.~\ref{Heff spin},\ref{Heff trans} below).

First we consider the effective low-energy model appropriate to the temperature range $(J,J')<T<U$, in which all dots become singly-occupied. Here the appropriate unity operator for the TQD Hilbert space is 
$\hat{1}_{\mathrm{so}}=\sum_{\sigma_1,\sigma_2,\sigma_3} |\sigma_1;\sigma_2;\sigma_3\rangle \langle\sigma_1;\sigma_2;\sigma_3|$, with $\sigma_i=\uparrow$$/$$\downarrow$ the spin of dot $i$. To second order in the dot-lead tunneling $V$, a spin-model of form $H_s=H_0+H_{\text{eff}}^s+H_{\text{tri}}^s$ 
arises; where $H_0$ describes the leads as above and:
\begin{equation}
\label{Hspin}
\begin{split}
H_{\text{tri}}^s = &J \hat{\textbf{S}}_2 \cdot (\hat{\textbf{S}}_1 + \hat{\textbf{S}}_3) + J'\hat{\textbf{S}}_1 \cdot \hat{\textbf{S}}_3\\
H_{\text{eff}}^s = &J_K (\hat{\textbf{S}}_1 \cdot \hat{\textbf{s}}_{L0} + \hat{\textbf{S}}_3\cdot\hat{\textbf{s}}_{R0})
\end{split}
\end{equation}
Here the effective exchange coupling is $\rho J_{K} =8\Gamma/(\pi U)$, with $\rho$ the lead density of states per orbital at the Fermi level; and $\Gamma =\pi V^{2}\rho_{T}^{\phantom\dagger}$ is the hybridization, with total lead density of states $\rho_{T}^{\phantom\dagger} =N\rho$, and $N$ the number of orbitals/$\textbf{k}$-states in the lead
(such that $NV^{2}$, and hence $\Gamma$, is finite in the continuum limit $N \rightarrow \infty$).
In Eq.~\ref{Hspin}, $\hat{\textbf{s}}_{\alpha 0}$ is the spin density of lead $\alpha=L(R)$ at dot $1(3)$ given by
\begin{subequations}\label{0 orb}
\begin{align}
\label{0 orb S}
\hat{\textbf{s}}_{\alpha 0}&=\sum_{\sigma,\sigma'} f^{\dagger}_{\alpha 0 \sigma} \boldsymbol{\sigma}_{\sigma\sigma'} f^{\phantom{\dagger}}_{\alpha 0 \sigma'}\\
\label{0 orb f}
f^{\dagger}_{\alpha 0 \sigma}&=\tfrac{1}{\sqrt{N}}\sum_{\textbf{k}} a^{\dagger}_{\alpha\textbf{k}\sigma}~,
\end{align}
\end{subequations}
with $\boldsymbol{\sigma}$ the Pauli matrices and $f^{\dagger}_{\alpha 0 \sigma}$ the creation operator for the `0'-orbital of the $\alpha=L,R$ Wilson chain.

As above, the lowest states of $H_{\text{tri}}^s$ are the doublets $|\pm;S^z\rangle$ given in Eq.~\ref{states}.
Provided they are not near-degenerate, only the lower doublet need be retained: $|+;S^z\rangle$ for $J'\gg J$ and $|-;S^z\rangle$ for $J'\ll J$. To first order in $J_{K}$, an effective low-energy model is then obtained simply by projecting into the reduced Hilbert space of the lowest doublet, using
\begin{equation}
\label{pm unity}
\hat{1}_{\pm}=\sum_{S^z} |\pm;S^z\rangle \langle\pm;S^z|
\end{equation}
for the appropriate $\gamma=+$ or $-$ doublet ground state. The resultant Hamiltonian
$H_{\text{eff},\gamma}=\hat{1}_{\gamma} H^{s}_{\text{eff}} \hat{1}_{\gamma}$ follows as
\begin{equation}
\label{Heff +-}
H_{\text{eff},\gamma}=\tfrac{1}{2}J_K \hat{1}_{\gamma}(\hat{\textbf{S}}_1+\hat{\textbf{S}}_3)\hat{1}_{\gamma}\cdot(\hat{\textbf{s}}_{L0} + \hat{\textbf{s}}_{R0})
\end{equation}
using the symmetry
$\hat{1}_{\gamma}\hat{\textbf{S}}_1\hat{1}_{\gamma}=\hat{1}_{\gamma}\hat{\textbf{S}}_3\hat{1}_{\gamma}$.

Eq.~\ref{Heff +-} is of two-channel Kondo form,
\begin{equation}
\label{Heff spin}
H_{\text{eff},\gamma}=J_{K\gamma} \hat{\textbf{S}}_{\gamma}\cdot(\hat{\textbf{s}}_{L0} + \hat{\textbf{s}}_{R0}),
\end{equation}
where $\hat{\textbf{S}}_{\gamma}$ is the spin-$\tfrac{1}{2}$ operator representing the appropriate doublet
$\gamma=+$ or $-$, with components $\hat{S}^{z}_{\gamma}=\sum_{S^{z}}|\gamma;S^{z}\rangle S^{z}\langle\gamma;S^{z}|$
and $\hat{S}^{\pm}_{\gamma} =|\gamma;\pm\tfrac{1}{2}\rangle\langle\gamma;\mp\tfrac{1}{2}|$. At this level of calculation the effective exchange coupling is given by
$J_{K\gamma}=\langle\gamma;+\tfrac{1}{2}|\hat{S}_1^z+\hat{S}_3^z|\gamma;+\tfrac{1}{2}\rangle J_K$; so from
Eq.~\ref{states} an AF effective Kondo coupling $J_{K-}=+\tfrac{2}{3}J_K$ then arises for the $|-;S^z\rangle$ ground state appropriate to $J'\ll J$, while for the $|+;S^z\rangle$  doublet (lowest for $J'\gg J$), $J_{K+}=0$ results. 
In the latter case, there is in fact a weak residual AF coupling: a full ${\cal{O}}(V^{2})$ Schrieffer-Wolff calculation gives precisely the two-channel Kondo model Eq. \ref{Heff spin} as one would expect, but with $J_{K\pm}$ given by
\begin{subequations}
\label{couplings}
\begin{align}
\rho J_{K+}=&\frac{4\Gamma}{\pi}\left\lbrace\frac{1}{3(J'-J)+2U}-\frac{1}{3J'+J+2U}\right\rbrace \\
\rho J_{K-}=&\frac{4\Gamma}{3\pi}\left\lbrace\frac{9}{(J-J')+2U}-\frac{1}{5J-J'+2U}\right\rbrace~;
\end{align}
\end{subequations}
yielding $\rho J_{K-} = \tfrac{2}{3}\rho J_{K} \propto \tfrac{\Gamma}{U}$ to leading  order in $1/U$, and \\ 
a much smaller but non-vanishing $\rho J_{K+} = \tfrac{4\Gamma J}{\pi U^{2}} +{\cal{O}}(\tfrac{1}{U^{3}})$
for the singlet-locked doublet $|+; S^{z}\rangle$, reflecting the residual AF coupling between the spin on dot 2 and the leads.

Hence, sufficiently deep in either regime $J'\ll J$ or $J'\gg J$, the low-energy behavior of the system is that of a 2CK model. The lowest spin-$\tfrac{1}{2}$ state of the TQD is thus overscreened by conduction electrons,  
embodied in the infrared 2CK fixed point describing the non-Fermi liquid ground 
state\cite{2CK:nozieres,2ck:cox_zaw,2ck:affleck2}, in which the partially quenched spin is characterised by a
residual entropy of $S_{\text{imp}}=\tfrac{1}{2}\ln 2$ ($k_B=1$); overscreening setting in below the characteristic two channel Kondo scale $T_{K,\gamma}^{2CK}$, determined from perturbative scaling as\cite{2CK:nozieres}
\begin{equation}
\label{pms}
T_{K,\gamma}^{2CK}\sim J_{K\gamma}\exp (-1/\rho J_{K\gamma}).
\end{equation}
Since $\hat{P}_{1,3}$ commutes with all components of $\hat{\textbf{S}}_{\gamma}$ in Eq.~\ref{Heff spin}, 
$[\hat{H}_{\text{eff},\gamma},\hat{P}_{1,3}]=0$, whence $1\leftrightarrow 3$ parity is conserved in the
effective low-energy model. Since that parity is determined by $\gamma = +$ or $-$, there are
two \emph{distinct} 2CK phases, which one thus expects to be separated by a quantum phase transition (QPT).

In the vicinity of the transition, \emph{ie} close to $J^{\prime} \simeq J$, neither of the two 2CK models in Eq. \ref{Heff spin} is of course sufficient to describe the low-energy physics: the states $|+;S^z\rangle$ and $|-;S^z\rangle$ are now near-degenerate, so both must be retained in the low-energy trimer manifold. 
Hence, defining $\hat{1}=\hat{1}_{+}+\hat{1}_{-}$ and proceeding in direct parallel to the discussion above,
an effective low-energy model in the vicinity of the transition is obtained from
$H_{\text{eff}}^{\text{trans}}=\hat{1}(H^{s}_{\text{eff}}+H^{s}_{\text{tri}})\hat{1}$. 
From Eq.~\ref{Hspin} for $H_{\text{eff}}^{s}$, using $\hat{\textbf{S}}_{1}=\hat{P}_{1,3}\hat{\textbf{S}}_{3}\hat{P}_{1,3}$ such that
$\langle\gamma ;S^{z}|\hat{\textbf{S}}_{1}|\gamma^{\prime} ;S^{z^{\prime}}\rangle =\gamma\gamma^{\prime}
\langle\gamma ;S^{z}|\hat{\textbf{S}}_{3}|\gamma^{\prime} ;S^{z^{\prime}}\rangle$ and hence
$\hat{1}_{\gamma}\hat{\textbf{S}}_{1}\hat{1}_{\gamma^{\prime}}=\gamma\gamma^{\prime}\hat{1}_{\gamma}\hat{\textbf{S}}_{3}\hat{1}_{\gamma^{\prime}}$, one obtains
\begin{equation}
\label{Heff tr}
\begin{split}
H_{\text{eff}}^{\text{trans}}=\sum_{\gamma=\pm}\tfrac{1}{2}J_K[&\hat{1}_{\gamma}(\hat{\textbf{S}}_1+\hat{\textbf{S}}_3)\hat{1}_{\gamma}\cdot(\hat{\textbf{s}}_{L0} + \hat{\textbf{s}}_{R0}) \\
\phantom{\sum_{\gamma=+,-}}+ &\hat{1}_{\gamma}(\hat{\textbf{S}}_1-\hat{\textbf{S}}_3)\hat{1}_{-\gamma}\cdot(\hat{\textbf{s}}_{L0} - \hat{\textbf{s}}_{R0})]\\
 + &\tfrac{1}{2}E_{\Delta}\left(\hat{1}_{+}-\hat{1}_{-}\right).
\end{split}
\end{equation}
The final term in Eq.~\ref{Heff tr} (arising from $\hat{1}H^{s}_{\text{tri}}\hat{1}$) is simply the energy difference between the $\gamma =+/-$ doublets, and  the first term is (see Eq.~\ref{Heff +-}) the 2CK coupling of \emph{each} doublet to the leads. It is helpful to recast Eq.~\ref{Heff tr} in terms of  spin-$\tfrac{1}{2}$ operators for real spin $\hat{\textbf{S}}$ ($=\sum_{\gamma}\hat{\textbf{S}}_{\gamma}$) and pseudospin $\hat{\boldsymbol{\mathcal{T}}}$ for the local Hilbert space, defined by
\begin{subequations}
\label{Sz,Tz}
\begin{align}
\hat{S}^z&= \sum_{\gamma =\pm,S^z}|\gamma;S^z\rangle S^z\langle\gamma;S^z|\\
\hat{\mathcal{T}}^z&= \sum_{\gamma =\pm,S^z}|\gamma;S^z\rangle \tfrac{1}{2}\gamma\langle\gamma;S^z|
\end{align}
\end{subequations}
and
\begin{subequations}
\label{Spm,Tpm}
\begin{align}
\hat{S}^{\pm}&= \sum_{\gamma}|\gamma;\pm\tfrac{1}{2}\rangle\langle\gamma;\mp\tfrac{1}{2}|\\
\hat{\mathcal{T}}^{\pm}&= \sum_{S^z}|\pm;S^z\rangle\langle\mp;S^z|~.
\end{align}
\end{subequations}
From Eq.~\ref{Sz,Tz}, the TQD doublets $|\gamma; S^{z}\rangle$ are each eigenstates of $\hat{S}^z$ and $\hat{\mathcal{T}}^z$; in particular, the eigenvalues of $\hat{\mathcal{T}}^z \equiv \tfrac{1}{2}(\hat{1}_{+}-\hat{1}_{-})$ correspond simply to (half) the parity of the appropriate doublet. 
By contrast, the doublets are interconverted by $\hat{\mathcal{T}}^{\pm}$ (Eq.~\ref{Spm,Tpm}b), 
$\hat{\mathcal{T}}^{+}|-;S^z\rangle=|+;S^z\rangle$ and $\hat{\mathcal{T}}^{-}|+;S^z\rangle=|-;S^z\rangle$,
acting to switch parity.

After simple if laborious algebra, Eq.~\ref{Heff tr} reduces to
\begin{equation}
\label{Heff trans}
\begin{split}
H_{\text{eff}}^{\text{trans}}&=(\tfrac{1}{2}J_A+J_B\hat{\mathcal{T}}^z)\hat{\textbf{S}}\cdot(\hat{\textbf{s}}_{L0} + \hat{\textbf{s}}_{R0})\\
&+J_{\text{mix}}(\hat{\mathcal{T}}^{+}+\hat{\mathcal{T}}^{-})\hat{\textbf{S}}\cdot(\hat{\textbf{s}}_{L0} - \hat{\textbf{s}}_{R0})\\
&+E_{\Delta}\hat{\mathcal{T}}^z,
\end{split}
\end{equation}
in terms of spin/pseudospin operators, with 
\begin{equation}
J_{A} = (J_{K+}+J_{K-}) ~~~~J_{B}=(J_{K+}-J_{K-})
\end{equation}
and $J_{\text{mix}}=\langle+;+\tfrac{1}{2}|\hat{S}_1^z-\hat{S}_3^z|-;+\tfrac{1}{2}\rangle J_K=\tfrac{1}{\sqrt{3}}J_K$ using Eq.~\ref{states}. [A full ${\cal{O}}(V^{2})$ Schrieffer-Wolff calculation again gives precisely the effective Hamiltonian Eq. \ref{Heff trans}, with 
\begin{equation}
\begin{split}
\rho J_{\text{mix}}=&\frac{2\Gamma}{\sqrt{3}\pi}\bigg{\lbrace} \frac{1}{3J'+J+2U}+\frac{1}{5J-J'+2U}\\
&\quad+\frac{3}{3(J'-J)+2U}+\frac{3}{(J-J')+2U}\bigg{\rbrace}
\end{split}
\end{equation}
recovering  $J_{\text{mix}} = \tfrac{1}{\sqrt{3}}J_{K}$ to leading order in $1/U$, and $J_{K\pm}$ given by Eq. \ref{couplings}.]

\begin{figure*}[t]
\includegraphics*[height=5cm]{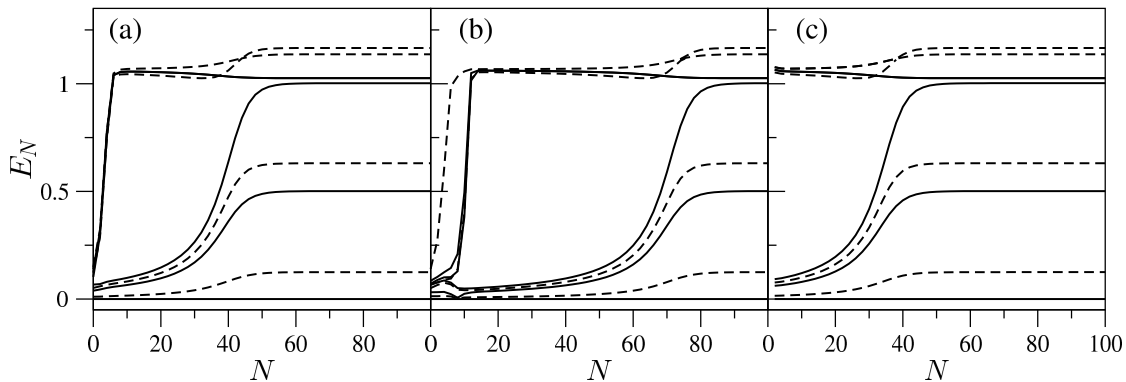}
\caption{\label{2ck levels}
Lowest five NRG energy levels of the left-charge/right-charge/spin subspaces $(Q_L, Q_R, S^z)\equiv (0,0,\tfrac{1}{2})$ [solid lines] and $(0,1,0)/(1,0,0)$ [dashed], versus (even) iteration number, $N$. 
Shown for $\rho J = 0.1$, $\tilde{U}=U/(\pi\Gamma)=7$, with $\rho J'=0$ [panel (a)] and $\rho J' = 0.105$ [(b)]. For comparison, (c) shows corresponding results for a pure (single-spin) 2CK model with $\rho J_K=0.055$.
}\end{figure*}

 Eq.~\ref{Heff trans} is the essential low-energy model applicable to the vicinity of the QPT; we study it directly via NRG in Sec.\ref{red}. The pseudospin operators can naturally be classified according to $1\leftrightarrow 3$ parity, the components of $\hat{\boldsymbol{\mathcal{T}}}$ having different parity under $\hat{P}_{1,3}$: using
$\hat{P}_{1,3}|\gamma ;S^{z}\rangle\langle\gamma^{\prime};S^{z}|=\gamma\gamma^{\prime}|\gamma ;S^{z}\rangle\langle\gamma^{\prime};S^{z}|\hat{P}_{1,3}$, Eqs.~\ref{Sz,Tz}b,\ref{Spm,Tpm}b
give $\hat{P}_{1,3}\hat{\mathcal{T}}^z=+\hat{\mathcal{T}}^z\hat{P}_{1,3}$, while
$\hat{P}_{1,3}\hat{\mathcal{T}}^{\pm}=-\hat{\mathcal{T}}^{\pm}\hat{P}_{1,3}$ for the raising/lowering components.
By contrast, all components of spin $\hat{\textbf{S}}$ (Eqs.~\ref{Sz,Tz}a,\ref{Spm,Tpm}a)
commute with $\hat{P}_{1,3}$. Hence, since the global (overall $L\leftrightarrow R$) parity must be conserved
($[H_{\text{eff}}^{\text{trans}}, \hat{P}_{L,R}]=0$), interactions involving the 
$\hat{\mathcal{T}}^{z}$ component of pseudospin can only couple to \emph{even} combinations 
$\hat{\textbf{s}}_{L0} + \hat{\textbf{s}}_{R0}$ of the lead spin densities (symmetric under
interchange of the $L/R$ lead labels), as in the first term of Eq. \ref{Heff trans} (or \ref{Heff tr});
while by the same token interactions involving $\hat{\mathcal{T}}^{\pm}$ must be associated with the
\emph{odd} (antisymmetric) combination $\hat{\textbf{s}}_{L0} - \hat{\textbf{s}}_{R0}$, as in the second term of 
Eq. \ref{Heff trans}. In the vicinity of the QPT the latter is of course the key interaction in $H_{\text{eff}}^{\text{trans}}$, since in switching the parity of the TQD states it in essence drives the transition between the two 2CK phases.

Finally, note that the last term of Eq.~\ref{Heff trans}, equivalent to a magnetic field acting on the pseudospin, energetically favors the $|-;S^z\rangle$ doublet ($\mathcal{T}^{z} =-\tfrac{1}{2}$) when their energy separation $E_{\Delta}>0$, and $|+;S^z\rangle$ ($\mathcal{T}^{z} =+\tfrac{1}{2}$) for $E_{\Delta}<0$. Hence, when
$|E_{\Delta}|$ is sufficiently large that only one of the doublets need be retained in the low-energy TQD manifold, the $\hat{\mathcal{T}}^{\pm}$ terms are obviously suppressed; and Eq.~\ref{Heff trans} then reduces as it must to one or other of the 2CK models Eq.~\ref{Heff spin}. In fact, as shown in Sec.~\ref{QPT} \emph{ff}, for any $|E_{\Delta}| \neq 0$, the $E_{\Delta}\hat{\mathcal{T}}^{z}$  term in Eq. \ref{Heff trans} ensures that one or other of the 2CK fixed points ultimately remains the stable low-temperature FP. 

By contrast, precisely at the point of frustration where
$|E_{\Delta}|=0$, the $\hat{\mathcal{T}}^{\pm}$ terms in Eq.~\ref{Heff trans} destabilize the 2CK FPs. The
resultant $H_{\text{eff}}^{\text{trans}}$ then describes the \emph{quantum critical point} which separates the
two 2CK phases, and which we consider in detail in Sec. \ref{QPT}.


\section{Properties of the 2CK Phases}\label{2CK}
The physical picture thus indicates that 2CK physics dominates the low-energy behavior of the model; with a QPT occurring as a function of $J'$ between two 2CK phases of distinct parity.

We now analyse the properties of each 2CK phase of the full model; using Wilson's NRG technique\cite{nrgreview,KWW}, employing a complete basis set of the Wilson chain\cite{asbasis} to calculate the full density matrix\cite{asbasis,fdmnrg} (for a recent review see [\onlinecite{nrg_rev}]). Calculations are typically performed for an NRG discretization parameter $\Lambda=3$, retaining the lowest $N_s=4000$ states per iteration. As above we choose for convenience $\epsilon = -\tfrac{1}{2}U$, and consider a symmetric constant density of states for each lead, with density of states per conduction orbital $\rho=1/(2D)$, and bandwidth $D/\Gamma=100\gg 1$ (such that results shown are essentially independent of $D$). In all calculations shown explicitly, we use a fixed
$\rho J=0.1$ and $\tilde{U}\equiv U/(\pi\Gamma)=7$, varying the exchange  $J'$ (see Fig.~\ref{dots})~\cite{note:jsimu}.

Fig.~\ref{2ck levels} shows the evolution of the lowest energy levels of the system as a function of NRG iteration
number $N$, exemplifying RG flow between different FPs of the model. Panel (a) is for a system deep in the $J'<J$ regime (specifically $\rho J'=0$), while panel (b) shows the energy levels for $\rho J'=0.105$ ($>\rho J$). For comparison, panel (c) is for a pure (single spin-$\tfrac{1}{2}$) 2CK model of form Eq.~\ref{Heff spin}, \emph{ie}
$H =J_{K} \hat{\textbf{S}}\cdot(\hat{\textbf{s}}_{L0} + \hat{\textbf{s}}_{R0})$,
with a Kondo coupling $\rho J_K=0.055$ chosen to be the same as the effective coupling $\rho J_{K-}$ of the ground state TQD doublet in (a) (as obtained from Eq.~\ref{couplings}(b)).  In both cases (a) and (b), the levels are seen to converge quite rapidly to their $N\rightarrow \infty$ values, which are clearly  those of the 2CK FP in Fig.~\ref{2ck levels}(c). These levels are of course characteristic of the 2CK FP, and --- after a trivial rescaling by a factor which depends on the NRG discretization parameter $\Lambda$ ---  are described by the fractions $0$, $\tfrac{1}{8}$, $\tfrac{1}{2}$, $\tfrac{5}{8}$, $1$... as determined from a conformal field theory analysis of the FP\cite{2ck:affleck1,2ck:ye}.

The iteration number by which the levels converge to the set of 2CK FP levels is however clearly different in (a) and (b), reflecting the different Kondo scales  characteristic of the two cases. Case (b) ($J^{\prime} >J$) flows close to the local moment (LM) FP between $N=10$ to $60$ (with levels naturally characteristic of the LM FP in that range), and approaches the stable 2CK FP by $N\simeq 80$. By contrast, convergence to the 2CK FP in (a) and (c) both occur at $N\approx 50$ (with a much shorter range of $N$ close to the LM FP). Since the iteration number is related to an effective temperature through~\cite{nrgreview,KWW}  $T/\Gamma\sim \Lambda^{-N/2}$, the 2CK Kondo scales of the three examples in Fig.~\ref{2ck levels} are thus exponentially small 
(as expected from Eq.~\ref{pms}); but with $T_{K}^{2CK}/\Gamma$ for (b) being some 8 orders of magnitude smaller than that of (a), reflecting the distinct nature of the coupling in the two cases, as discussed in Sec.~\ref{eff models}.


\begin{figure}[t]
\includegraphics*[height=9.5cm]{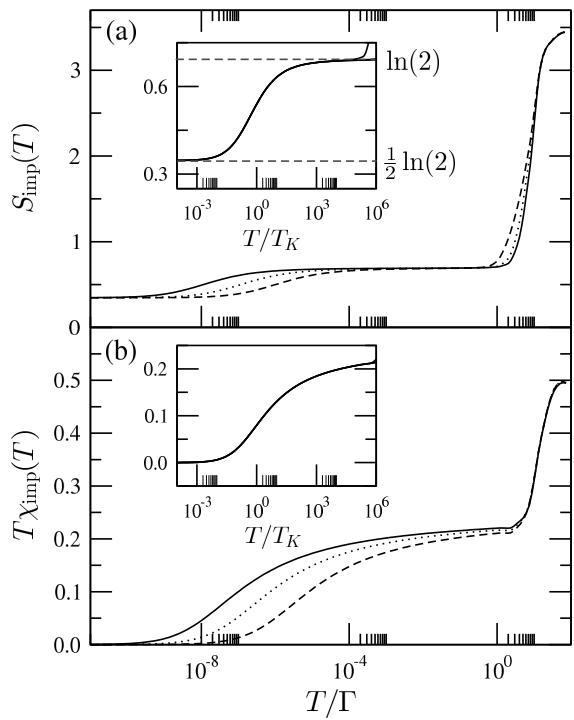}
\caption{\label{thermo J'<J}
Thermodynamics of the $J'<J$ phase. Shown for fixed $\tilde{U}=7$ and $\rho J=0.1$, varying $\rho J'=0$ (solid lines), $0.0375$ (dotted) and $0.075$ (dashed). Panel (a) shows $S_{\text{imp}}(T)$, (b) shows 
$T\chi_{\text{imp}}(T)$, both \emph{vs} $T/\Gamma$. Insets show the universality arising on rescaling in terms of the 2CK Kondo temperature, $T_K$, all curves collapsing to a common form.
}\end{figure}

\subsection{Thermodynamics}\label{thermo 2CK}

NRG results for thermodynamics in each phase are now considered. We focus on the `impurity' (TQD) contribution\cite{KWW,nrg_rev}
to the entropy, $S_{\text{imp}}(T)$, and the uniform spin susceptibility, $\chi_{\text{imp}}(T)
=\langle (\hat{S}^{z})^{2}\rangle_{\mathrm{imp}}/T$ (here $\hat{S}^{z}$ refers to the spin of the entire system
and $\langle\hat{\Omega}\rangle_{\mathrm{imp}} =\langle\hat{\Omega}\rangle - \langle\hat{\Omega}\rangle_{0}$
with $\langle\hat{\Omega}\rangle_{0}$ denoting a thermal average in the absence of the TQD);
the temperature ($T$) dependences of which provide clear signatures of the underlying FPs reached under renormalization on  progressive reduction of the temperature/energy scale\cite{KWW,nrg_rev}. 

\subsubsection{$J'<J$}\label{TD J'<J}
For the $J'\ll J$ regime the effective 2CK model Eq.~\ref{Heff spin} should describe the system for $T\ll |J-J'|=|E_{\Delta}|$, where the lowest TQD doublet, in this case the odd parity state  $|-;S^z\rangle$, couples 
symmetrically to the leads. 2CK physics is thus expected below $T\sim T_K \equiv T_{K-}^{2CK}$.

For the full model, Fig.~\ref{thermo J'<J} shows both $S_{\text{imp}}(T)$ [panel (a)] and $T\chi_{\text{imp}}(T)$ [panel (b)] for fixed $\tilde{U}=7$ and $\rho J=0.1$, with $J'/J=0$, $0.375$ and $0.75$, all deep in the $J'<J$ regime. In each case, at high temperatures $T\gtrsim U$ the behavior is governed by the free orbital~\cite{KWW} (FO) FP, with all possible states of the TQD thermally accessible, and hence $S_{\text{imp}}=\ln(32)$.
For $T\lesssim U$ the dots become in essence singly occupied, and hence an entropy of $\ln(8)$ is 
expected~\cite{note:ln8}. Below $T\sim |E_{\Delta}|$ all but the lowest trimer doublet 
is projected out. Thus $S_{\text{imp}}$ approaches $\ln(2)$, signifying the LM FP where the lowest doublet is 
essentially a free spin-$\tfrac{1}{2}$ disconnected from the leads (Eq.~\ref{Heff spin} with $J_{K-}\equiv 0$).

\begin{figure}[t]
\includegraphics*[height=5.5cm]{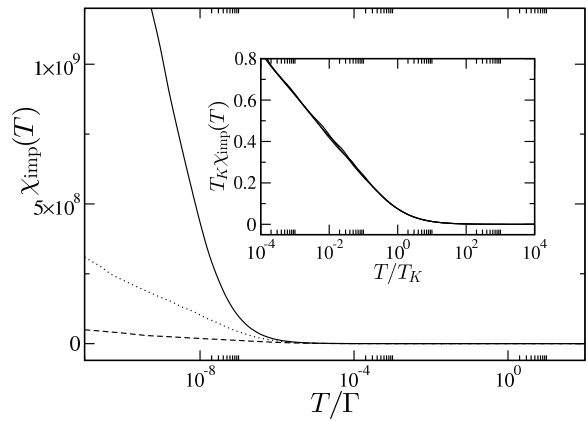}
\caption{\label{chi J'<J}
Spin susceptibility $\chi_{\text{imp}}(T)$ \emph{v}s $T/\Gamma$ in the $J'<J$ regime, for same parameters as Fig.~\ref{thermo J'<J}. Inset: scaling collapse onto the universal form $T_{K} \chi_{\text{imp}}(T)= \alpha\ln(T_K/T)$.
}\end{figure}

The LM FP is however unstable, and the system always flows to the stable 2CK FP below $T \sim T_{K}$, recovering
(Fig.~\ref{thermo J'<J}(a)) the $\tfrac{1}{2}\ln 2$ residual entropy known to be characteristic of the 2CK FP\cite{TD:Destri,TD:Tsvelick,TD:Sacramento}.
In practice we may define a Kondo temperature $T_{K} \equiv T_{K}^{S}$ through the entropy, via
$S_{\text{imp}}(T^S_K)=0.55$ (suitably between $\ln 2$ and $\tfrac{1}{2}\ln 2$); or alternatively
$T_{K} \equiv T_{K}^{\chi}$ through the spin susceptibility
via $T^{\chi}_K\chi_{\text{imp}}(T^{\chi}_K)=0.07$ (as in  [\onlinecite{KWW}]). Deep in the 2CK phases  the two definitions are of course equivalent ($T_K^{\chi}\equiv T_K^S\equiv T_{K}$), probing as they do the common characteristic scale associated with flow to the 2CK FP.
The inset to Fig.~\ref{thermo J'<J}(a) shows the entropy of the three systems rescaled in terms of $T/T_K$;
showing scaling collapse to a common functional form, \emph{ie} the universality characteristic of the
crossover from the LM FP to the stable 2CK FP\cite{TD:Sacramento}.

The underlying FPs of the model are likewise evident from $T\chi_{\text{imp}}(T)$, Fig.~\ref{thermo J'<J}(b). The highest $T$ behavior corresponds to two uncorrelated sites (dots 1 and 3 of the TQD) and a free spin (dot 2). Hence $T\chi_{\text{imp}}(T)=2\times\tfrac{1}{8}+\tfrac{1}{4}=\tfrac{1}{2}$, readily understood as the mean $\langle(S^z)^2\rangle$ of the quasidegenerate states. On decreasing $T$ the LM FP is again rapidly reached, the lowest TQD doublet following the free spin-$\tfrac{1}{2}$ Curie law $T\chi_{\text{imp}}(T)=\tfrac{1}{4}$. Below $T_K$ the spin susceptibility is quenched\cite{TD:Destri,2ck:affleck1,TD:Sacramento} in the sense that $T\chi_{\text{imp}}(T)\rightarrow 0$ as $T\rightarrow 0$. In the inset to Fig.~\ref{thermo J'<J}(b)
the data are rescaled in terms of $T/T_{K}$, again showing universality in the approach to the 2CK FP\cite{TD:Sacramento}.

The characteristic low-$T$ logarithmic divergence\cite{TD:Destri,2ck:affleck1,TD:Sacramento} of $\chi_{\text{imp}}(T)$ itself is evident in Fig.~\ref{chi J'<J} for the same parameters as Fig.~\ref{thermo J'<J}. The slopes of the divergence vary widely between the three cases, but all collapse to the universal form\cite{TD:Sacramento} $T_K \chi_{\text{imp}}(T)= \alpha\ln(T_K/T)$ (with $\alpha$ a constant), as seen in the inset of the figure.

\begin{figure}[t]
\includegraphics*[height=9.5cm]{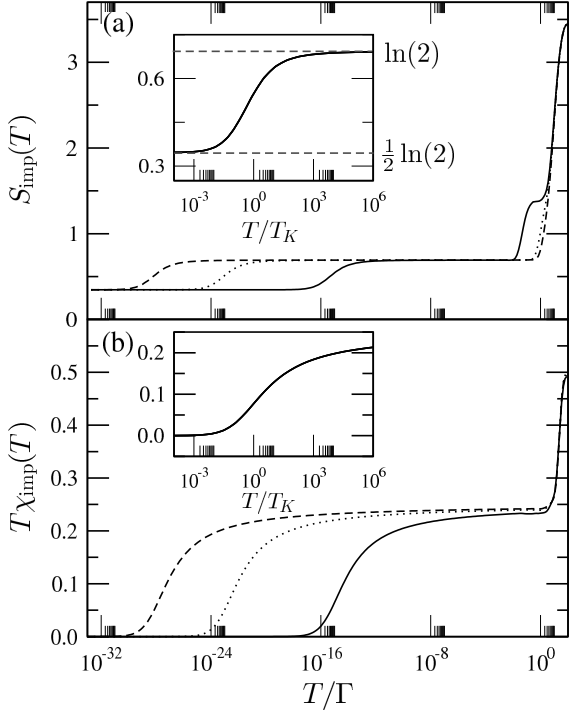}
\caption{\label{thermo J'>J}
Thermodynamics of the $J'>J$ phase. Shown for fixed $\tilde{U}=7$, $\rho J=0.1$, varying $\rho J'=0.105$ (solid lines), $0.115$ (dotted) and $0.125$ (dashed). Panel (a) again shows $S_{\text{imp}}(T)$, (b) shows $T\chi_{\text{imp}}(T)$, both \emph{v}s $T/\Gamma$. Insets show rescaling in terms of the Kondo temperature, $T_K$, yielding the same universal 2CK scaling curves as in Fig.~\ref{thermo J'<J}.
}\end{figure}

\subsubsection{$J'>J$}\label{TD J'>J}
From the arguments of Sec.~\ref{eff models}, 2CK physics is again expected at low-energies in the 
$J^{\prime} \gg J$ regime, where the even-parity  $|+;S^z\rangle$ doublet is now the lowest TQD state.
This is confirmed in Fig.~\ref{thermo J'>J}, where (in direct parallel to Fig.~\ref{thermo J'<J} for $J'\ll J$) thermodynamics are shown for fixed $\tilde{U}=7$ and $\rho J=0.1$, but now with $J'/J=1.05$, $1.15$ and $1.25$: the scaling forms for $S_{\mathrm{imp}}(T)$ and $T\chi_{\mathrm{imp}}(T)$ \emph{vs} $T/T_{K}$, shown in the insets of Fig.~\ref{thermo J'>J}, are precisely those arising for $J' \ll J$ and shown in Fig.~\ref{thermo J'<J}.

  While the low-energy physics in the two regimes $J'\ll J$ and $J' \gg J$ have common 2CK scaling behavior, 
we add that the Kondo scales $T_{K}$ themselves evolve very differently with $J^{\prime}$; as will be considered explicitly in Sec.~\ref{Kondo scales} (see Fig.~\ref{Tk scales}), and evident even from cursory comparison of Figs.~\ref{thermo J'>J},\ref{thermo J'<J}.


\subsection{Dynamics}
\label{dyn 2CK}
We turn now to dynamics, focussing largely on the local single-particle spectrum of dot 1 (or, equivalently by symmetry, dot 3): $D_1(\omega)=-\tfrac{1}{\pi}\text{Im}G_1(\omega)$, with $G_1(\omega)$ the local retarded Green function. We obtain it through the Dyson equation,
\begin{equation}
\label{dyson}
[G_1(\omega)]^{-1}=[G_1^0(\omega)]^{-1}-\Sigma_1(\omega),
\end{equation}
where $G_1^0(\omega)$ is the non-interacting propagator (\emph{ie} for $U=0=J=J'$); and $\Sigma_1(\omega)$ is
the proper electron self-energy, with $\Sigma_1(\omega)\equiv \Sigma_1^R(\omega) - i\Sigma_1^I(\omega)$ (such that
$\Sigma_1^I(\omega) \geq 0$). The non-interacting $G_1^0(\omega)$ is given trivially by\cite{hewson}
$[G_1^0(\omega)]^{-1}=\omega^{+}-\epsilon-\Gamma(\omega)$ (with $\omega^{+}=\omega+i0^{+}$), where $\Gamma(\omega) = \sum_{\textbf{k}}V^{2}[\omega^{+}-\epsilon_{\textbf{k}}]^{-1}$ is the usual one-electron hybridization function for coupling of dot 1 to the left lead: $\Gamma(\omega)\equiv \Gamma^R(\omega)-i\Gamma^I(\omega)$, with $\Gamma^{I}(\omega) =\Gamma$ ($=\pi V^{2}\rho_{T}^{\phantom\dagger}$) for
all $|\omega| < D$ inside the band, and $\Gamma^R(\omega =0)=0$.

An expression for the electron self-energy $\Sigma_{1}(\omega)$ is also readily obtained using equation of motion methods\cite{hewson,EOM}. It is given by
\begin{equation}
\label{self-energy}
\begin{split}
\Sigma_{1}(\omega)=[G_{1}(\omega)]^{-1}\bigg{\{}U&\langle\langle c^{\phantom{\dagger}}_{1\uparrow}\hat{n}^{\phantom{\dagger}}_{1\downarrow};c^{\dagger}_{1\uparrow} \rangle\rangle_{\omega}^{\phantom\dagger}\\
+\tfrac{1}{2}J&\langle\langle c^{\phantom{\dagger}}_{1\downarrow} \hat{S}_2^{-}+c^{\phantom{\dagger}}_{1\uparrow} \hat{S}_2^{z};c^{\dagger}_{1\uparrow} \rangle\rangle_{\omega}^{\phantom\dagger}\\
+\tfrac{1}{2}J'&\langle\langle c^{\phantom{\dagger}}_{1\downarrow} \hat{S}_3^{-}+c^{\phantom{\dagger}}_{1\uparrow} \hat{S}_3^{z};c^{\dagger}_{1\uparrow} \rangle\rangle_{\omega}^{\phantom\dagger} \bigg{\}}
\end{split}
\end{equation}
where $\langle\langle \hat{A};\hat{B} \rangle\rangle_{\omega}^{\phantom\dagger}$ is the Fourier transform of the retarded correlator $\langle\langle \hat{A}(t_1);\hat{B}(t_2) \rangle\rangle=-i\theta(t_1-t_2)\langle\{\hat{A}(t_1),\hat{B}(t_2)\}\rangle$, and where the local Green function 
itself is given by $G_{1}(\omega)=\langle\langle c^{\phantom{\dagger}}_{1\sigma};c^{\dagger}_{1\sigma} \rangle\rangle_{\omega}^{\phantom\dagger}$ (independent of spin $\sigma$ in the absence of a magnetic field). The self-energy can be calculated directly within the density matrix formulation of the NRG\cite{asbasis,fdmnrg,UFG,nrg_rev}, via Eq.~\ref{self-energy}; with $G_{1}(\omega)$ then obtained from Eq.~\ref{dyson}. $D_1(\omega)$ calculated in this way is highly accurate\cite{UFG,nrg_rev}, and automatically guarantees correct normalization of the spectrum\cite{asbasis,fdmnrg}.

To motivate study of the spectrum $D_{1}$ ($\equiv D_{1}(\omega;T)$), we note that it controls the zero-bias conductance through dot 1. The TQD does not of course mediate current from the $L$ to $R$ leads, since the internal couplings between constituent dots are pure exchange (Fig.~\ref{dots} and Eq.~\ref{Hfull}). However the $L$ and $R$ leads in Fig.~\ref{dots} can obviously each be `split' in two (symmetrically, to preserve overall $L\leftrightarrow R$ symmetry), enabling a current to be driven through dot 1 or dot 3; with the same zero-bias conductance in either case, by symmetry. Following Meir and Wingreen\cite{wingreenspec}, the resultant conductance follows as
\begin{equation}
\label{zbc}
\frac{G_c(T)}{G_0}=\int_{-\infty}^{\infty} -\frac{\partial f(\omega)}{\partial \omega}~~ \pi \Gamma D_1(\omega;T) ~\text{d}\omega 
\end{equation}
where $f(\omega)=[e^{\omega /T}+1]^{-1}$ is the Fermi function (with $\omega =0$ the Fermi level) and $G_{0} = 2e^{2}/h$ the conductance quantum. The ($\omega,T$)-dependence of $D_{1}$ thus controls the conductance, and for $T=0$ in particular $G_c(T=0)/G_0~=~\pi\Gamma D_1(\omega =0;T=0)$.  From Eq.~\ref{dyson}, the 
local propagator for $\omega =0 =T$ may be expressed as $[G_1(\omega=0)]^{-1}=-\epsilon^{*}+i\Gamma^{*}$ in terms of the renormalized single-particle level $\epsilon^{*}$ and renormalized hybridization $\Gamma^{*}$, given  by
\begin{subequations}
\begin{align}\label{renorm1}
\epsilon^{*}&=\epsilon+\Sigma_1^R(0)\\
\Gamma^{*}&=\Gamma+\Sigma_1^I(0)
\label{renorm1b}
\end{align}
\end{subequations}
in terms of the $\omega =0$ value of the self-energy at $T=0$; and hence from Eq.~\ref{zbc}:
\begin{equation}
\label{zbcT=0renorm}
\frac{G_{c}(T=0)}{G_{0}}~=~\pi\Gamma D_{1}(0;0)~=~\frac{^{\phantom *}\Gamma^{\phantom *}}{^{\phantom *}\Gamma^{*}}~\frac{1}{1+\left(\frac{\epsilon^{*}}{\Gamma^{*}}\right)^{2}}  
\end{equation}

  NRG results for single-particle dynamics are considered below, but first a question arises.
We have argued above that, sufficiently deep in either regime $J'<J$ or $J'>J$, the low-energy physics of the full
three-site TQD model must reduce to that of a single spin-$\tfrac{1}{2}$ 2CK model (of form Eq.~\ref{Heff spin}). The question is: to which dynamical property of a pure 2CK model should the spectral density $\pi\Gamma D_{1}(\omega)$ be compared? To answer this, note first that Eq.~\ref{zbc}  may be written equivalently as
\begin{equation}
\label{zbctmatrix}
\frac{G_c(T)}{G_0}=\int_{-\infty}^{\infty} -\frac{\partial f(\omega)}{\partial \omega}~\left[ -\pi\rho_{T}^{\phantom\dagger}\text{Im}~t_{L}(\omega)\right] ~\text{d}\omega 
\end{equation}
in terms of the t-matrix, $t_{L}(\omega)$, for the $L$ lead; with $t_{\alpha}(\omega)$ defined in the usual way in terms of scattering of electron states in the $\alpha =L,R$ lead, via
\begin{equation}
\label{t-matrix}
G_{\alpha\textbf{k},\alpha\textbf{k}'}(\omega)=\frac{\delta_{\textbf{kk}'}}{\omega^{+}-\epsilon_{\textbf{k}}}+\frac{1}{\omega^{+}-\epsilon_{\textbf{k}}} ~t_{\alpha}(\omega)~\frac{1}{\omega^{+}-\epsilon_{\textbf{k}'}}
\end{equation}
where $G_{\alpha\textbf{k},\alpha\textbf{k}'}(\omega)=\langle\langle a^{\phantom{\dagger}}_{\alpha\textbf{k}\sigma};a^{\dagger}_{\alpha\textbf{k}'\sigma}\rangle\rangle_{\omega}$ 
is the propagator for the lead states. Using equation of motion methods~\cite{hewson,EOM} it is straightforward to show that $t_{L}(\omega)=V^{2}G_{1}(\omega)$ (likewise $t_{R}(\omega)=V^{2}G_{3}(\omega) \equiv t_{L}(\omega)$);
so $-\pi\rho_{T}^{\phantom\dagger}\text{Im}t_{L}(\omega)=\pi\Gamma D_{1}(\omega;T)$ (recall $\Gamma =\pi V^{2}\rho_{T}^{\phantom\dagger}$), hence the equivalence of Eqs.~\ref{zbc},\ref{zbctmatrix}. 

To compare $\pi\Gamma D_{1}(\omega;T)$ for the full TQD model to results for a single-spin 2CK model \cite{2ck:affleck2,2ck:expt_propose1,2ck:dyn_johannesson,2ck:dyn_anders,2ck:propose1,2ck:dyn_toth1},
we thus require $t_{L}(\omega)\equiv t_{L}^{2CK}(\omega)$ for the latter.  Using the definition of the 0-orbital of a Wilson chain (Eq.~\ref{0 orb f}), the Hamiltonian for a single-spin 2CK model is
\begin{equation}
\label{2CK H}
\begin{split}
H_{\text{2CK}}=&\sum_{\alpha\textbf{k}\sigma}\epsilon_{\textbf{k}}a^{\dagger}_{\alpha\textbf{k}\sigma} a^{\phantom{\dagger}}_{\alpha\textbf{k}\sigma}\\
& +\sum_{\alpha\textbf{kk}'}\tfrac{J_K}{2N}\big{[}\hat{S^z}(a^{\dagger}_{\alpha\textbf{k}\uparrow} a^{\phantom{\dagger}}_{\alpha\textbf{k}'\uparrow}-a^{\dagger}_{\alpha\textbf{k}\downarrow} a^{\phantom{\dagger}}_{\alpha\textbf{k}'\downarrow})\\
&\quad\qquad+\hat{S}^{+}a^{\dagger}_{\alpha\textbf{k}\downarrow} a^{\phantom{\dagger}}_{\alpha\textbf{k}'\uparrow}+\hat{S}^{-}a^{\dagger}_{\alpha\textbf{k}\uparrow} a^{\phantom{\dagger}}_{\alpha\textbf{k}'\downarrow}\big{]};
\end{split}
\end{equation}
and for this case equations of motion again yield Eq.~\ref{t-matrix} for 
$G_{\alpha\textbf{k},\alpha\textbf{k}'}(\omega)$, but now with $t_{L}(\omega) = \frac{J_{K}^{2}}{4N}G_{s}(\omega)$
where
\begin{equation}
\label{Gs}
G_{s}(\omega)~=~\langle\langle \hat{S}^{-} f^{\phantom{\dagger}}_{L0\downarrow}+\hat{S}^z f^{\phantom{\dagger}}_{L0\uparrow}; \hat{S}^{+} f^{\dagger}_{L0\downarrow}+\hat{S}^z f^{\dagger}_{L0\uparrow} \rangle\rangle_{\omega}^{\phantom\dagger}~.
\end{equation}
Comparison of  $-\pi\rho_{T}^{\phantom\dagger}\text{Im}t_{L}(\omega)$ for the full TQD model with its pure 2CK counterpart (using $\rho_{T}^{\phantom\dagger}/N=\rho = 1/(2D)$) then gives the desired correspondence 
\begin{equation}
\pi\Gamma D_{1}(\omega)~~\leftrightarrow ~~D~\rho_{K}^{\phantom\dagger}(\omega)
\end{equation}
with spectral density
\begin{equation}
\label{rhoK}
\rho_{K}^{\phantom\dagger}(\omega)~=~ -\tfrac{\pi}{2}(\rho J_{K})^{2}~\text{Im} G_{s}(\omega)
\end{equation}
for the single-spin 2CK model~\cite{note:Gs}.


\subsubsection{Dynamics: results}
\label{dynamicsresults}

\begin{figure}[t]
\includegraphics*[height=5.5cm]{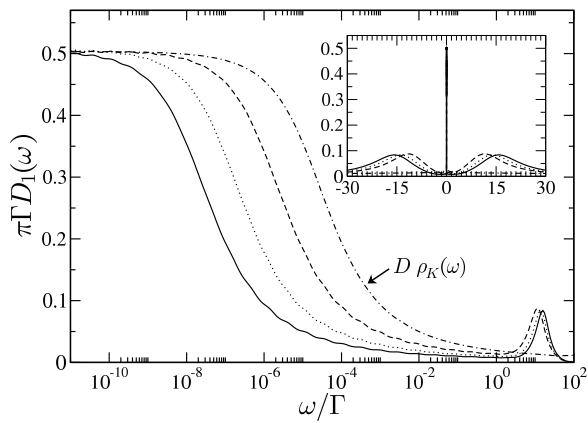}
\caption{\label{Dw 2CK}
T=0 local spectrum $\pi\Gamma D_1(\omega)$ \emph{vs} $\omega/\Gamma$ in the $J'<J$ regime. Shown 
(as in Fig.~\ref{thermo J'<J}) for fixed $\tilde{U}=7$ and $\rho J =0.1$, varying
$J'/J = 0$ (solid line), $0.375$ (dotted) and $0.75$ (dashed). The dot-dashed line is the spectral density 
$D\rho_{K}^{}(\omega)$ (Eq.~\ref{rhoK}) for a single-spin 2CK model (Eq.~\ref{2CK H}), with $\rho J_K=0.07$. Inset: results shown on a linear frequency scale.
}
\end{figure}

Fig.~\ref{Dw 2CK} shows the $T=0$ spectrum $\pi\Gamma D_1(\omega)$ \emph{vs} $\omega/\Gamma$, in the $J'<J$ regime (for the same bare parameters as Fig.~\ref{thermo J'<J} for thermodynamics). The main panel shows results on a log scale for $\omega >0$; and particle-hole symmetry for $\epsilon=-\tfrac{1}{2}U$ means $D_1(\omega)= D_1(-\omega)$ (as seen in the inset). The figure also shows $D\rho_K(\omega)^{\phantom\dagger}$ for the single-spin 2CK model
(for $\rho J_K=0.07$, ensuring an exponentially small Kondo scale but otherwise chosen arbitrarily).

We first comment on the high-energy spectral features (`Hubbard satellites') in $D_{1}(\omega)$. As usual, these reflect simple one-electron addition to the isolated TQD ground state. The lowest such excitation from the 
$|-;S^z\rangle$ TQD ground state incurs an energy cost $\Delta E_{-}=\tfrac{1}{4}(J-J')-\epsilon$.
On tunnel coupling to the leads these features are naturally broadened, but the satellites are centered 
on $\omega\simeq \Delta E_{-}$, indeed seen from  Fig.~\ref{Dw 2CK} to shift slightly to lower $\omega$ on increasing $J'$. In the $J'>J$ regime (not shown), directly analogous behavior arises.  Here the $|+;S^z\rangle$ doublet is the TQD ground state, the Hubbard satellites in $D_{1}(\omega)$ are now centered around $\omega\simeq \Delta E_{+}=\tfrac{3}{4}(J'-J)-\epsilon$, and hence shift to \emph{higher} frequency as $J'$ is increased.
By contrast, see Fig.~\ref{Dw 2CK}, Hubbard satellites are simply absent in 
$\rho_{K}^{\phantom\dagger}(\omega)$ -- addition/removal excitations are  
suppressed by construction in modeling a dot strictly as a spin.

 The most important characteristic of the spectra in Fig.~\ref{Dw 2CK} is of course the low-energy Kondo resonance,
 the form of which reflects RG flow in the vicinity of the stable 2CK fixed point. At the Fermi level $\omega =0$ in particular, $\pi\Gamma D_{1}(\omega=0) =\tfrac{1}{2}$ in all cases (likewise  $D\rho_{K}^{\phantom\dagger}(0)=\tfrac{1}{2}$) -- \emph{ie} reaches \emph{half} the unitarity limit, a hallmark of the 2CK FP, likewise known from study of the two-channel (quadrupolar) single-impurity Anderson model~\cite{2ck:dyn_anders,2ck:dyn_johannesson}.
  
\begin{figure}[t]
\includegraphics*[height=5.5cm]{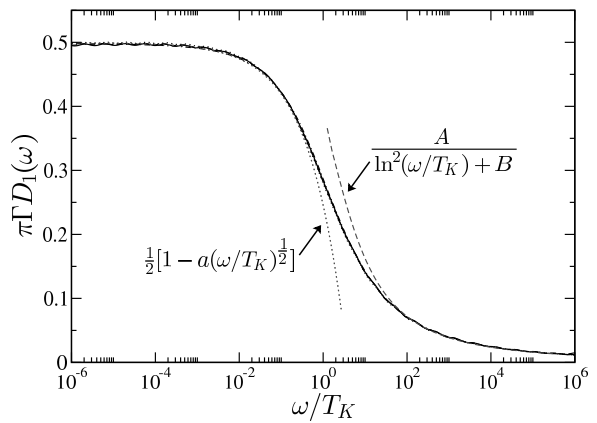}
\caption{\label{Dw scaling}
$T=0$ scaling spectrum for the 2CK phases. $\pi\Gamma D_1(\omega)$ \emph{vs} $\omega/T_K$ is shown both
in the $J'<J$ regime (for the bare parameter sets used in Fig.~\ref{thermo J'<J}) \emph{and} for the $J'>J$ regime (with the parameters used in Fig.~\ref{thermo J'>J}).  $D\rho_K(\omega)$ \emph{vs} $\omega/T_K$ for the single-spin 2CK model is also shown. All spectra collapse to the same universal form. The dotted line shows the low-$|\omega|/T_{K}$ asymptotic behavior $\pi\Gamma D_1(\omega)= \tfrac{1}{2}[1-a(|\omega|/T_K)^{\tfrac{1}{2}}]$, 
while the dashed line describes the high-$|\omega|/T_{K}$ scaling behavior, $\pi\Gamma D_1(\omega)= A/[\ln^{2}(|\omega|/T_K)+B]$.
}\end{figure}

   Universal scaling of single-particle dynamics is considered in Fig.~\ref{Dw scaling}, where
$\pi\Gamma D_1(\omega)$ \emph{vs} $\omega/T_K$ is shown (with $T_K$ defined in practice from thermodynamics, as in Sec.~\ref{thermo 2CK}). The figure includes the three examples shown in Fig.~\ref{Dw 2CK} for the 
$J'<J$ (odd parity ground state) regime, as well as spectra for the $J' >J$ (even parity) regime for the
same parameters as Fig.~\ref{thermo J'>J}; together with $D\rho_K(\omega)^{\phantom\dagger}$ \emph{vs} $\omega/T_{K}$ for the pure 2CK model.
As seen from Fig.~\ref{Dw scaling}, all spectra collapse
perfectly to a universal scaling form; confirming that the TQD model -- be it in the $J' >J$ or
$J' <J$ regime -- is described by the 2CK model at low-energies.

  As shown in Fig.~\ref{Dw scaling}, the leading low-frequency asymptotics of the scaling spectrum are found to be 
\begin{equation}
\label{lowfreqD}
\pi\Gamma D_{1}(\omega)~\overset{\tfrac{|\omega|}{T_{K}}\ll 1}\sim ~
\tfrac{1}{2}[1-a\left(|\omega|/T_K\right)^{\tfrac{1}{2}}]
\end{equation}
with the constant $a \simeq 0.47$; as consistent with that found for the quadrupolar Anderson impurity model~\cite{2ck:dyn_anders}, and in  contrast to the $[1-a(\omega/T_K)^2]$ decay characteristic of the normal Fermi liquid FP in single-channel models. For $|\omega|/T_{K} \gg 1$ by contrast, as also shown in Fig.~\ref{Dw scaling}, the asymptotic behavior is 
\begin{equation}
\label{highfreqD}
\pi\Gamma D_{1}(\omega)~\overset{\tfrac{|\omega|}{T_{K}}\gg 1}\sim ~
A/[\ln^{2}(|\omega|/T_K)+B]
\end{equation}
(with $A$ and $B$ pure constants ${\cal{O}}(1)$); which form is also asymptotically common to the single-channel
spin-$\tfrac{1}{2}$ Anderson model~\cite{NLDscaling} and its $SU(2N)$ generalization~\cite{LMA:SU(2N)}.
This is physically natural, the origin of the `high' energy leading $\sim 1/\ln^{2}(|\omega|/T_{K})$ logarithms
being spin-flip scattering~\cite{hewson}, which occurs for both single- and two-channel models.

\begin{figure}[t]
\includegraphics*[height=10cm]{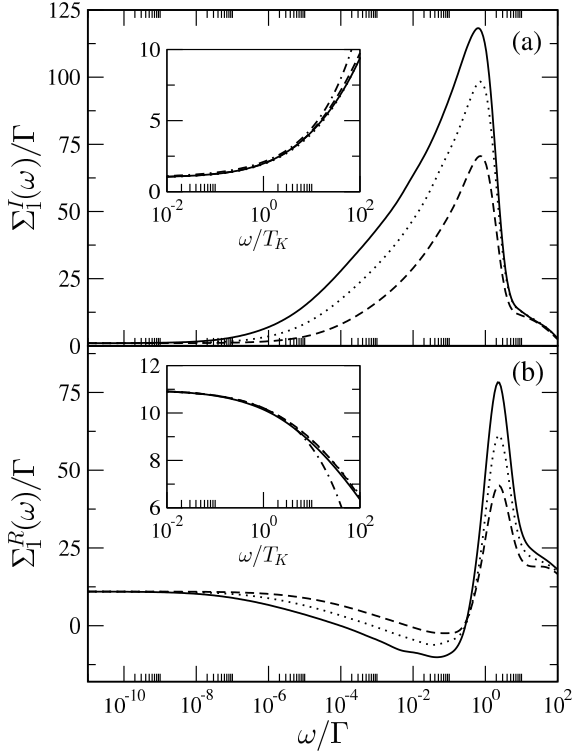}
\caption{\label{SE}
Electron self-energy 
$\Sigma_1(\omega)/\Gamma$ 
\emph{vs} $\omega/\Gamma$ for the same parameters as 
Fig.~\ref{Dw 2CK}.
Panel (a): imaginary part, $\Sigma_{1}^{I}(\omega)/\Gamma$; panel (b): real part, $\Sigma_{1}^{R}(\omega)/\Gamma$. Insets: universal scaling forms as a function of $\omega/T_K$, with the dot-dashed lines showing the low-frequency asymptotic behavior $\Sigma_1^I(\omega)/\Gamma =1+c|\omega/T_K|^{\tfrac{1}{2}}$, and  $\Sigma_1^R(\omega)/\Gamma =-\epsilon/\Gamma-\text{sgn}(\omega)c^{\prime}|\omega/T_K|^{\tfrac{1}{2}}$.
}
\end{figure}

We consider now the electron self-energy of dot 1 in the full TQD model, the real and imaginary parts of which are shown in Fig.~\ref{SE} and naturally exhibit scaling in terms of $\omega/T_{K}$. For $\Sigma_1^I(\omega)$ we find the low-$\omega/T_{K}$ asymptotic behavior to be $\Sigma_1^I(\omega)/\Gamma = 1+c|\omega/T_K^{2CK}|^{\tfrac{1}{2}}$ as shown explicitly in the inset to panel (a). At the Fermi level in particular, $\Sigma_1^I(\omega=0)=\Gamma$, in marked contrast to a Fermi liquid phase for which $\Sigma^I(\omega=0)=0$ generically; and leading to a renormalized hybridization (Eq.~\ref{renorm1b}) $\Gamma^{*}=\Gamma + \Sigma_1^I(\omega=0) = 2\Gamma$. The real part of the self-energy is shown in panel (b), and at the Fermi level in particular is found to be
$\Sigma_1^R(\omega=0)=-\epsilon$, such that the renormalized level $\epsilon^{*} =0$ (Eq.~\ref{renorm1}).

  Since $\Gamma^{*}=2\Gamma$ and $\epsilon^{*}=0$, it follows from Eq.~\ref{zbcT=0renorm} that the $T=0$
zero-bias conductance for the TQD model reduces to one-half the unitarity limit, \emph{ie} $G_{c}(T=0)/G_{0}=\tfrac{1}{2}$, as known to be the case for a single-spin 2CK 
model~\cite{2ck:affleck2,2ck:expt_propose1,2ck:propose1,2ck:dyn_toth1}. Moreover, although the explicit case for which we have shown results is $\epsilon =-U/2$ where the TQD model is particle-hole symmetric, we note that
$\Gamma^{*}=2\Gamma$ and $\epsilon_{*}=0$
 -- and hence $\pi\Gamma D_{1}(\omega) =\tfrac{1}{2}$ and
$G_{c}(0)/G_{0} =\tfrac{1}{2}$ -- 
is also found to hold robustly on moving away from particle-hole symmetry.

  The results above refer to dynamics for $T=0$, and we now touch on $T>0$. Universality implies
$\pi\Gamma D_{1}(\omega;T)$ depends on $\omega/T_{K}$ and $T/T_{K}$. Without loss of generality 
it may be cast as
\begin{equation}
\label{scalingfreqT}
\pi\Gamma D_{1}(\omega;T)=\pi\Gamma D_{1}(0;T) -\frac{a}{2}\left(\frac{|\omega|}{T_{K}}\right)^{\tfrac{1}{2}}
~g\left(\frac{\omega}{T_{K}};\frac{T}{T_{K}}\right)
\end{equation}
where the scaling function $g$ satisfies  $g(0;0) =1$ (such that Eq.~\ref{scalingfreqT} for $T=0$ reduces asymptotically to
Eq.~\ref{lowfreqD} for $\pi\Gamma D_{1}(\omega;T=0)$). For the Fermi level $\omega =0$ in particular, we find from NRG that the leading low-$T/T_{K}$ dependence of $\pi\Gamma D_{1}(0;T)$ is
\begin{equation}
\label{scalingfreq0}
\pi\Gamma D_{1}(0;T)~\overset{\tfrac{T}{T_{K}}\ll 1}\sim ~
\tfrac{1}{2}[1-b\left(T/T_K\right)^{\tfrac{1}{2}}]
\end{equation}
with $b \simeq a$ a constant. Employing Eq.~\ref{scalingfreqT} in Eq.~\ref{zbc} for $G_{c}(T)$, and rescaling the resultant non-trivial integral (arising from the second term of Eq.~\ref{scalingfreqT}) by exploiting the fact that the Fermi function depends solely on $\omega/T$, gives the leading low-$T/T_{K}$ behavior of the zero-bias conductance as
\begin{equation}
\label{zbcfiniteT}
\frac{G_{c}(T)}{G_{0}}~=~\tfrac{1}{2}[ 1-\gamma \left(T/T_{K}\right)^{\tfrac{1}{2}}]
\end{equation}
where $\gamma$ is a constant ($\gamma = b+a^{\prime}$ with $a^{\prime} =a\sqrt{\pi}\eta (\tfrac{1}{2}) \simeq 1.07 a$
and $\eta(z)$ the Dirichlet $\eta$-function). And the behavior Eq.~\ref{zbcfiniteT} is precisely that known to arise for the single-spin 2CK model~\cite{2ck:affleck2,2ck:expt_propose1,2ck:propose1,2ck:dyn_toth1}.

\begin{figure}[t]
\includegraphics*[height=6cm]{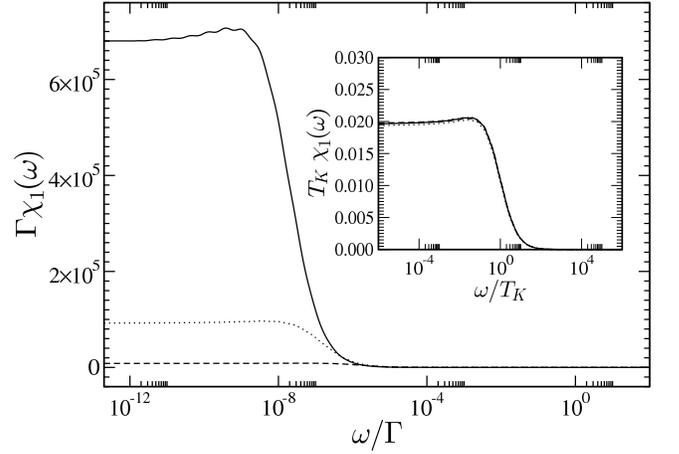}
\caption{\label{dyn susc}
Local dynamic susceptibility $\Gamma\chi_{1}^{}(\omega)$ \emph{vs} $\omega/\Gamma$ at $T=0$ for the same bare parameters as Fig.~\ref{Dw 2CK}. Inset: showing universal scaling of  $T_{K}\chi_{1}^{}(\omega)$ \emph{vs} $\omega/T_{K}$.
}\end{figure}

Finally, we consider briefly the local dynamic spin susceptibility of dot 1 in the 2CK phases, given by
$\chi_{1}^{}(\omega)=-\tfrac{1}{\pi}\text{Im}\langle\langle \hat{S}_1^z;\hat{S}_1^z \rangle \rangle_{\omega}$.
Fig.~\ref{dyn susc} shows  $\chi_{1}^{}(\omega)$ \emph{vs} $\omega/\Gamma$ (for $T=0$) in the $J'<J$ regime, with the same bare parameter sets as Fig.~\ref{Dw 2CK} (the behavior discussed below being applicable in both 2CK phases). In the standard single-channel Anderson model, $\chi_{1}^{}(\omega)$ exhibits characteristic
low-$\omega$ Fermi liquid behavior~\cite{shiba,hewson},  $\chi_{1}^{}(\omega)\propto \omega$.
By contrast, it is known~\cite{2ck:dyn_anders,2ck:dyn_bradley} that at the 2CK FP, $\chi_{1}^{}(0)$ plateaus at a finite constant; itself proportional to the slope of the log divergence  of $\chi_{\text{imp}}(T)$ as $T\rightarrow 0$. In Fig.~\ref{chi J'<J} (inset) for the full TQD model we showed  the uniform static spin susceptibility to have a
low-$T$ log divergence of form $\chi_{\text{imp}}(T)= \tfrac{\alpha}{T_{K}}\ln(T_K/T)$, with slope $\alpha/T_{K}$.
Thus we also expect $T_{K}\chi_{1}^{}(\omega)$ to exhibit universality as a function of $\omega/T_{K}$; as indeed confirmed in the inset of Fig.~\ref{dyn susc}.

\subsection{Kondo Scales}\label{Kondo scales}
\begin{figure}[t]
\includegraphics*[height=6cm]{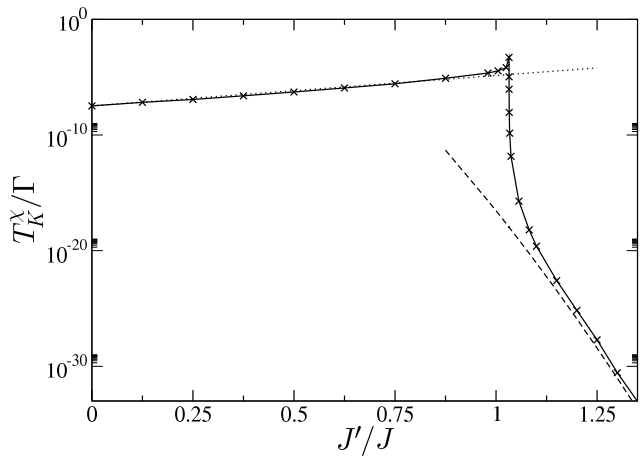}
\caption{\label{Tk scales}
Kondo temperature $T_K^{\chi}/\Gamma$ (points, with solid line as guide to eye) \emph{vs} $J'/J$ for fixed $\tilde{U} =7$ and $\rho J =0.1$; as defined (Sec.~\ref{thermo 2CK}) via $T_K^{\chi}\chi_{\text{imp}}^{}(T_K^{\chi})=0.07$. Kondo scales obtained  from perturbative scaling  
(Eq.~\ref{pms}) are also shown, with $\rho J_{K+}$ (for $J'/J>1$) from Eq.~\ref{couplings}(a) (dashed line) and $\rho J_{K-}$ (for $J'/J<1$) from Eq.~\ref{couplings}(b) (dotted); with a trivial constant prefactor  adjusted to fit the data.
}\end{figure}

The energy scale on which 2CK physics emerges in the full TQD model is of course the 2-channel
Kondo temperature. $T_{K}$ itself varies markedly with the bare model parameters; with qualitatively different behavior for $J'/J<1$ and $J'/J>1$ that reflects the distinct nature (Sec.~\ref{eff models}) of the lowest-energy
TQD doublets in the two regimes, $|-;S^{z}\rangle$ and $|+;S^{z}\rangle$ respectively.
This is shown in Fig.~\ref{Tk scales}, considering explicitly the
Kondo scale $T_{K} \equiv T_{K}^{\chi}$ obtained (as in Sec.~\ref{thermo 2CK}) from the uniform susceptibility,
and showing $T_{K}^{\chi}/\Gamma$ \emph{vs} $J'/J$. In the $J'<J$ regime a relatively modest increase in $T_K$ occurs on increasing $J'/J$; while for $J'>J$ by contrast $T_K$ diminishes very rapidly, reflecting physically the weak residual AF coupling (Sec.~\ref{eff models}) between the spin on dot 2 and the leads,  when the singlet-locked $|+;S^{z}\rangle$ doublet is the TQD ground state.

In Sec.~\ref{eff models} we showed that the full TQD $H$ maps onto an effective single-spin 2CK model 
(Eq.~\ref{Heff spin}) in each of the $J'<J$ and $>J$ regimes, provided the separation between the TQD doublets
$|E_{\Delta}|=|J-J'|$ is sufficiently large that one or other alone dominates the low-energy physics. The resultant 
Kondo temperatures for the two regimes are then given from perturbative scaling by Eqs.~\ref{pms},\ref{couplings}.
These are compared directly to the NRG results for $T_{K}^{\chi}$ in Fig.~\ref{Tk scales}, and are seen to
agree quantitatively for $|E_{\Delta}|/J\gtrsim 0.1$, \emph{ie} throughout the great majority of each of the two 2CK regimes.

\section{Quantum phase transition}
\label{QPT}
When $|E_{\Delta}| =|J-J'|$ is sufficiently small however, neither TQD doublet alone dominates the low-energy physics, both must then be included in the low-energy trimer manifold, and as shown in Sec.~\ref{eff models} the low-energy behavior is no longer described by an effective single-spin 2CK model, but rather by Eq.~\ref{Heff trans}; with coupling between doublets of distinct parity embodied in the pseudospin raising/lowering terms.
We show in the following that these additional terms in the effective Hamiltonian drive a quantum phase transition between the two 2CK phases, occurring at the point of inherent magnetic frustration.

\begin{figure}[t]
\includegraphics*[height=6cm]{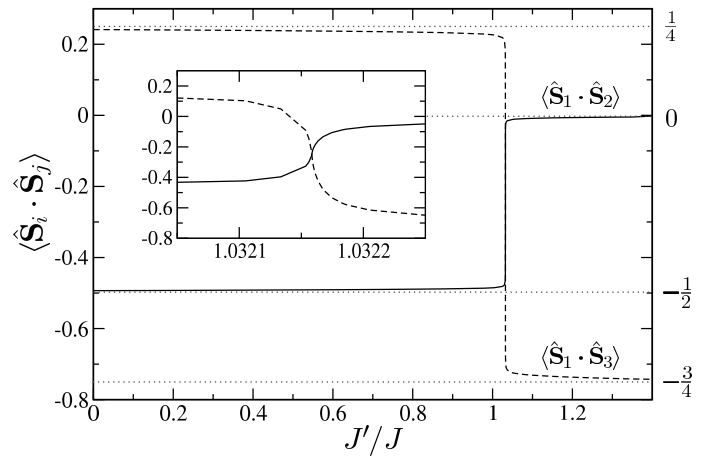}
\caption{\label{sisj trans}
$T=0$ spin-spin correlation functions $\langle\hat{\textbf{S}}_1 \cdot \hat{\textbf{S}}_2 \rangle$ (solid line) and  $\langle\hat{\textbf{S}}_1 \cdot \hat{\textbf{S}}_3 \rangle$ (dashed) \emph{vs} $J'/J$, for fixed $\tilde{U}=7$ and $\rho J=0.1$. Inset: magnification around the point of frustration,
where $\langle\hat{\textbf{S}}_1 \cdot \hat{\textbf{S}}_2 \rangle=\langle\hat{\textbf{S}}_1 \cdot \hat{\textbf{S}}_3 \rangle$, occurring at $J'/J=1.032..$.
}\end{figure}

  In the \emph{isolated} TQD, frustration occurs at $J'=J$ where the TQD doublets are degenerate, a trivial level-crossing `transition' occurring in this case as $J'_{c}=J$ is crossed. As a result, ground state properties
in general change discontinuously across $J'_{c}$; exemplified \emph{eg} by the spin correlation functions
$\langle\hat{\textbf{S}}_1 \cdot \hat{\textbf{S}}_2 \rangle$ and
$\langle\hat{\textbf{S}}_1 \cdot \hat{\textbf{S}}_3 \rangle$ which (as in Sec.~\ref{TQD}) change abruptly at $J'_{c}$ from $-\tfrac{1}{2}$ and $+\tfrac{1}{4}$ respectively in the ($-$)-parity phase
$J'<J'_{c}$ to $0$ and $-\tfrac{3}{4}$ respectively in the ($+$)-parity phase for $J'>J'_{c}$.
That situation changes qualitatively on coupling to the leads, as illustrated in Fig.~\ref{sisj trans}
where the $T=0$ spin correlators  $\langle\hat{\textbf{S}}_1 \cdot \hat{\textbf{S}}_2 \rangle$ 
and $\langle\hat{\textbf{S}}_1 \cdot \hat{\textbf{S}}_3 \rangle$ are shown as a function of $J'/J$ for the full model~\cite{zitkoTQD2ch}.  While their behavior sufficiently deep in either of the 2CK regimes naturally accords with expectations from the isolated TQD, they now vary continuously with $J'/J$; as seen clearly from the magnification of the crossover region in the inset. The value of $J'/J$ for which ($\langle\hat{\textbf{S}}_3 \cdot \hat{\textbf{S}}_2 \rangle \equiv$) $\langle\hat{\textbf{S}}_1 \cdot \hat{\textbf{S}}_2 \rangle=\langle\hat{\textbf{S}}_1 \cdot \hat{\textbf{S}}_3 \rangle$ is the point of complete magnetic frustration, and as seen from the inset to Fig.~\ref{sisj trans} is naturally slightly renormalized from unity, to
$J_c'/J = 1.03216...$ in the example shown.

We shall see in the following sections that the QPT occurs at $\tilde{E}_{\Delta}=J'_c-J'=0$, with a quantum critical point separating the two 2CK phases of distinct parity.

\subsection{Physical picture of the transition}
\label{PhysPicQPT}
Before showing NRG results, we give heuristic physical arguments for the behavior of the system in the vicinity of the QPT, based on the effective low-energy Hamiltonian $H_{\text{eff}}^{\text{trans}}$ (Eq.~\ref{Heff trans}) in which both TQD doublet states are retained in the low-energy trimer manifold.  $H_{\text{eff}}^{\text{trans}}$ is 
cast in terms of spin-$\tfrac{1}{2}$ operators for both real spin ($\hat{\textbf{S}}$) and pseudospin ($\hat{\boldsymbol{\mathcal{T}}}$) for the local TQD Hilbert space, the distinct parity of the two doublets being reflected in the $\mathcal{T}^z =\pm \tfrac{1}{2}$ components of pseudospin. 

In  Eq.~\ref{Heff trans} the exchange couplings $\tfrac{1}{2}(J_{A}\pm J_{B})$ $(\equiv J_{K\pm})$ 
are both $>0$ (AF), in which case it is readily argued (and confirmed explicitly by NRG) that the low-energy FP structure of $H_{\text{eff}}^{\text{trans}}$ is independent of $J_{B}$. For simplicity in the following we thus consider $J_{B}=0$ in Eq.~\ref{Heff trans}, \emph{ie}
\begin{equation}
\label{Hefftrans1}
\begin{split}
H_{\text{eff}}^{\text{trans}}&=\tfrac{1}{2}J_A \hat{\textbf{S}}\cdot(\hat{\textbf{s}}_{L0} + \hat{\textbf{s}}_{R0})\\
&+J_{\text{mix}}(\hat{\mathcal{T}}^{+}+\hat{\mathcal{T}}^{-})\hat{\textbf{S}}\cdot(\hat{\textbf{s}}_{L0} - \hat{\textbf{s}}_{R0})\\
&+\tilde{E}_{\Delta}\hat{\mathcal{T}}^z,
\end{split}
\end{equation}
with $\tilde{E}_{\Delta} = J'_{c}-J'$ (rather than the `bare' $E_{\Delta} = J-J'$, allowing simply for 
the slight renormalization of $J'_{c}$ away from the value $J'_{c}=J$ in the isolated TQD, as above).

 For $\tilde{E}_{\Delta} \neq 0$, a `magnetic field' acts on the $z$-component of pseudospin, and
as mentioned in Sec.~\ref{eff models} the low-energy physics is ultimately that of a single-spin 2CK FP (we return to it below). But now consider $\tilde{E}_{\Delta} = 0$ in Eq.~\ref{Hefftrans1}; \emph{ie} $J'=J'_{c}$, corresponding to the transition itself. In this case no field acts on the pseudospin, its `$z$-component' as such being arbitrary. Recalling that $\hat{\mathcal{T}}^{+}+\hat{\mathcal{T}}^{-} = 2\hat{\mathcal{T}}^x$ in Eq.~\ref{Hefftrans1}, and then performing a trivial rotation of the pseudospin axes from $(x,y,z) \rightarrow (z,x,y)$, Eq.~\ref{Hefftrans1} may be written as:
\begin{equation}
\label{Hefftrans2}
\begin{split}
H_{\text{eff}}^{\text{trans}}(\tilde{E}_{\Delta}=0)&=(\tfrac{1}{2}J_A +2J_\text{mix}\hat{\mathcal{T}}^z) \hat{\textbf{S}}\cdot \hat{\textbf{s}}_{L0} \\
&+(\tfrac{1}{2}J_A -2J_\text{mix}\hat{\mathcal{T}}^z)\hat{\textbf{S}}\cdot \hat{\textbf{s}}_{R0}
\end{split}
\end{equation}
This Hamiltonian commutes with $\hat{\mathcal{T}}^z$, so is strictly separable into disjoint
$\mathcal{T}^z =+\tfrac{1}{2}$ and $-\tfrac{1}{2}$ sectors, \emph{ie} the pseudospin is free.
But in either given $\mathcal{T}^z$ sector, the spin $\hat{\textbf{S}}$ clearly couples asymmetrically to the $L$ and $R$ channels, \emph{ie} one has \emph{channel anisotropy}. In the single spin-$\tfrac{1}{2}$ 2CK model, channel anisotropy is of course well known~\cite{2CK:nozieres,TD:Sacramento,2ck:affleck1,2ck:cox_zaw,AndreiJerez2CK} to destabilise the 2CK FP: the spin-$\tfrac{1}{2}$ is instead fully quenched by the channel to which it is most strongly AF exchange-coupled (and is decoupled from the second conduction channel/lead). The ultimate stable fixed point is then the \emph{single channel} strong coupling (SC) FP, reached below a temperature scale $T_{K}^{1CK} \propto \exp(-1/\rho J_{>})$ with $J_{>}$ the larger of the $L/R$ exchange couplings.

  In the present context, Eq.~\ref{Hefftrans2}, the situation is then clear: in the $\mathcal{T}^z =+\tfrac{1}{2}$
sector the spin is wholly quenched by coupling to the $L$ lead/channel ($J_{\text{mix}}>0$), while for
$\mathcal{T}^z =-\tfrac{1}{2}$ it is quenched by coupling to the $R$ lead; the  temperature scale for quenching
in either case being the one-channel Kondo scale $T_{K}^{1CK} \propto \exp(-1/\rho J_{>})$ with $J_{>}\equiv \tfrac{1}{2}J_{A}+J_{\text{mix}}$. 

 The stable FP is clearly a doubled version of the SC FP, the `doubling' reflecting the free pseudospin;
with an associated $T=0$ residual entropy of $S_{\text{imp}} = \ln 2$, and a finite uniform spin susceptibility $\chi^{}_{\text{imp}}(0) \propto 1/T_{K}^{1CK}$ symptomatic of the 1-channel quenched spin-$\tfrac{1}{2}$. Since this FP is distinct from the stable 2CK FP arising for $\tilde{E}_{\Delta} \neq 0$ away from the QPT, it corresponds to a critical FP (CFP); with Eq.~\ref{Hefftrans2} the effective Hamiltonian for the quantum critical point (QCP) itself. 

 The `entire' $T$-dependence of thermodynamics at the QCP is also readily inferred, since the pseudospin is ubiquitously free and spin-quenching is characterised by the single, finite scale $T_{K}^{1CK}$.
For $T \gg T_{K}^{1CK}$ one expects  $S_{\text{imp}}(T) =\ln 4$ (reflecting the free pseudospin 
\emph{and} the free spin), with a free spin-$\tfrac{1}{2}$ Curie law $T\chi^{}_{\text{imp}}(T) \sim \tfrac{1}{4}$;
the `$SU(2) \times SU(2)$' FP Hamiltonian here being Eq.~\ref{Hefftrans2} with all exchange couplings set to zero. $T \sim T_{K}^{1CK}$ then sets the scale for spin quenching and the crossover to the CFP discussed above. Hence, at the QCP, the $T$-dependence of the entropy should be given by $S_{\text{imp}}(T)=\ln 2 +S_{\text{imp}}^{1CK}(T)$, with $S_{\text{imp}}^{1CK}(T)$ the entropy for  a 1-channel Kondo model with Kondo scale $T_{K}^{1CK}$; 
likewise the uniform spin susceptibility should be $\chi^{}_{\text{imp}}(T) =\chi^{1CK}_{\text{imp}}(T)$.

  Having considered the QCP we return briefly to a non-vanishing 
$|\tilde{E}_{\Delta}| \ll T_{K}^{1CK}$, small compared to the 1CK scale.
In this case we expect flow to the CFP to be cut off at a characteristic 2CK scale, $T \sim T_{K}$,
below which the system flows to the stable 2CK FP; and with $T_{K} \equiv T_{K}(\tilde{E}_{\Delta})$
vanishing as $|\tilde{E}_{\Delta}| \rightarrow 0$, indicative of the QPT. We examine this via NRG in the following section, but note here that the considerations above would lead us to expect a crossover in $S_{\text{imp}}(T)$ from $\ln 4$ to $\ln2$  on the scale $T \sim T_{K}^{1CK}$, followed by a crossover for $T \sim T_{K}$ to the $\tfrac{1}{2}\ln 2$ characteristic of the stable 2CK FP. Likewise for $\chi^{}_{\text{imp}}(T)$
we anticipate the  Curie law $\chi^{}_{\text{imp}}(T) \sim 1/4T$ for $T \gg T_{K}^{1CK}$, increasing to
$\chi^{}_{\text{imp}}(T) \propto 1/T_{K}^{1CK}$ for $T \sim T_{K}^{1CK}$, before crossing over to the  divergent 2CK behavior $\chi^{}_{\text{imp}}(T) \propto \ln (T_{K}/T)$ for $T \lesssim T_{K}$.

  Finally, for a larger non-zero $|\tilde{E}_{\Delta}| \gg T_{K}^{1CK}$, no RG flow in the vicinity of the CFP is expected, and the $T_{K}^{1CK}$ scale is then irrelevant: below $T \sim |\tilde{E}_{\Delta}|\gg T_{K}^{1CK}$ one or other of the pseudospin $z$-components is simply frozen out, $S_{\text{imp}}(T)$ crossing over from
$\ln 4$ to $\ln 2$ (here indicative of a frozen pseudospin but a free spin-$\tfrac{1}{2}$), and 
then to the stable 2CK FP value $\tfrac{1}{2}\ln 2$ on the 2CK scale $T \sim T_{K}$.
The physics here is simply that of a single spin-$\tfrac{1}{2}$ 2CK model, as already considered in the
preceding sections.


\begin{figure*}[t]
\includegraphics*[height=5cm]{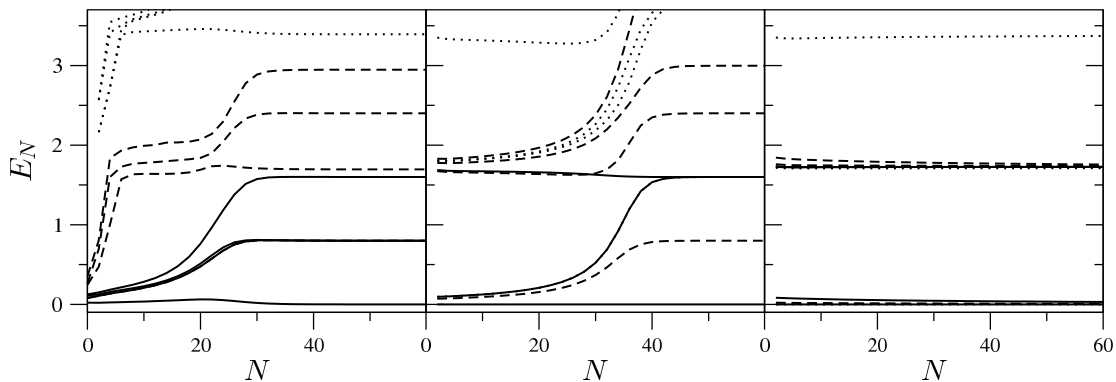}
\caption{\label{levels QCP}
NRG energy levels for even iteration number, $N$. The lowest 4 levels are shown in each charge/spin subspace for: the full TQD model precisely at the transition [\emph{left panel}], a 1CK model with $\rho J_K=+0.056$ [\emph{middle}] and $\rho J_K=- 0.056$ [\emph{right}]. For the full model, subspaces shown are $(Q_L, Q_R, S^z)\equiv (0,1,0)/(1,0,0)$ [solid lines], $(0,1,1)/(1,0,1)$ [dashed], and $(0,3,0)/(3,0,0)$ [dotted]. For the 1CK models, the subspaces are $(Q, S^z)\equiv (0,0)$ [solid lines], $(1,\tfrac{1}{2})$ [dashed], and $(0,1)$ [dotted].
}\end{figure*}

\subsection{Critical fixed point: NRG}
\label{nature of qcp}

The physical arguments given above imply that the NRG energy levels associated with the critical FP itself, should consist of \emph{both} a set of single-channel strong coupling FP levels, reflecting  spin-quenching to one lead (as arises for an AF-coupled single-channel spin-$\tfrac{1}{2}$ Kondo model); \emph{and} a set of levels for a free conduction band, reflecting decoupling of the spin from the other lead (as arises at the trivial weak coupling FP for a ferromagnetically coupled single-channel spin-$\tfrac{1}{2}$ Kondo model~\cite{hewson}).

  Before proceeding we simply demonstrate that this is indeed the case. The left panel of Fig.~\ref{levels QCP}
shows the lowest NRG energy levels of the full TQD model at the transition ($J'=J'_{c}$), as a function of iteration number, $N$. The CFP levels are well converged by $N=40$ (flow to the CFP from the $SU(2)\times SU(2)$ FP beginning for $N \approx 26$).
For comparison we also show the lowest NRG levels for a representative 1CK model with AF coupling (middle panel) and with ferromagnetic coupling (right panel). The CFP level structure can indeed be seen to comprise \emph{both} sets of strong coupling (middle panel) and weak coupling (right panel) levels. Indeed for the full TQD 
model, levels in the subspaces $(Q_L, Q_R, S^z)\equiv (x,y,S^z)$ and $(y,x,S^z)$ are degenerate, 
so  overall $L/R$ symmetry is naturally unbroken, and the strong coupling and weak coupling levels form symmetrically in each channel.


\subsection{Thermodynamics of the transition}\label{trans thermo}
We turn now to thermodynamics in the vicinity of the transition, obtained via NRG for the full TQD model; considering temperatures $T \ll \Gamma$ (\emph{ie} without comment on non-universal high-temperatures, where the same behavior as in Sec.~\ref{thermo 2CK} naturally occurs).

\begin{figure}[t]
\includegraphics*[height=5.75cm]{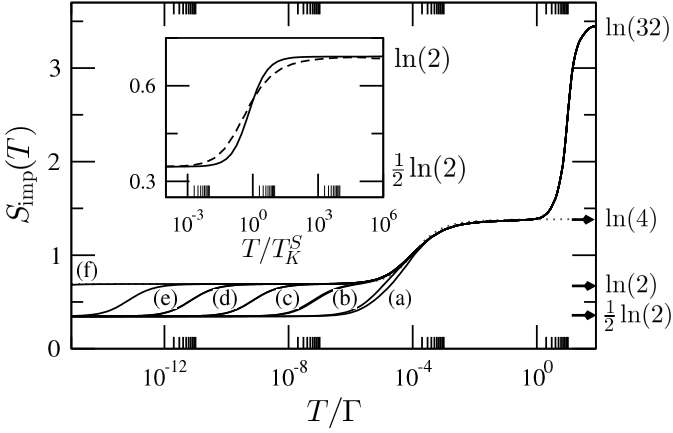}
\caption{\label{td trans}
Entropy  $S_{\text{imp}}(T)$ \emph{vs} $T/\Gamma$ on progressively approaching the QPT as: $\tilde{E}_{\Delta}=(J'_c-J')=\pm\lambda T_K^{c}$, with $T_K^{c}/\Gamma=10^{-4}$ (as discussed in text) and results shown for $\lambda=10^n$ with $n=0,-1,-2,-3,-4$ (lines (a)-(e) respectively). The QCP itself ($\lambda =0$) is shown as line (f), while the dotted line shows $\ln 2+S_{\text{imp}}^{1CK}(T)$ calculated for a single-channel Kondo model with $T_{K}^{1CK} =T_{K}^{c}$; the two coinciding for all $T/\Gamma \lesssim 1$. Inset: results in (b)-(e) rescaled (solid line) in terms of the vanishing  Kondo scale $T_{K}\equiv T_{K}^{S}$, showing universal crossover from the CFP to the stable 2CK FP; and contrasted to that arising in the single-spin 2CK model (dashed line).
}\end{figure}

Fig.~\ref{td trans} shows the $T/\Gamma$-dependence of the entropy $S_{\text{imp}}(T)$. Results are given for fixed 
$\tilde{U}=7$ and $\rho J=0.1$, varying $J'$ as the transition --- occurring at $J'=J'_{c}$ determined as above from the spin correlators  --- is approached from either side, according to $\tilde{E}_{\Delta} =(J'_c-J')=\pm\lambda  T_K^{c}$; with results shown for $\lambda=10^n$ and $n=0,-1,-2,-3,-4$ [lines (a-e) in Fig.~\ref{td trans}], chosen to approach progressively the transition, itself occurring at $\lambda=0$ [shown as line (f)].
Here, $T_K^{c}/\Gamma=10^{-4}$ is the value, at $J'=J'_c$, of the Kondo scale $T_K^{\chi}$ obtained from $T\chi^{}_{\text{imp}}(T)$ as in Sec.~\ref{thermo 2CK}; and we will in fact shortly identify $T_{K}^{c}$ as corresponding to the effective single-channel Kondo scale $T_{K}^{1CK}$ discussed in Sec.~\ref{PhysPicQPT} and associated with the QCP. Lines (a) in Fig.~\ref{td trans} are thus for $J'$ being $\pm T_{K}^{c}$ away from the transition; while (b-e) show the behavior for $|J'_{c}-J'| \ll T_{K}^{c}$ as the QPT is approached.

$S_{\text{imp}}(T)$ self-evidently shows a QPT: one sees clearly a low-energy scale
$T_{K} \equiv T_{K}^{S}$ in the vicinity of the transition, reflected in the crossover to the stable 
2CK FP value $S_{\text{imp}} =\tfrac{1}{2}\ln 2$ (and determined in practice via $S_{\text{imp}}(T^S_K)=0.55$ as in Sec.~\ref{thermo 2CK}); with $T_{K}$ vanishing precisely at the transition itself (line (f)).

\begin{figure}[t]
\includegraphics*[height=6cm]{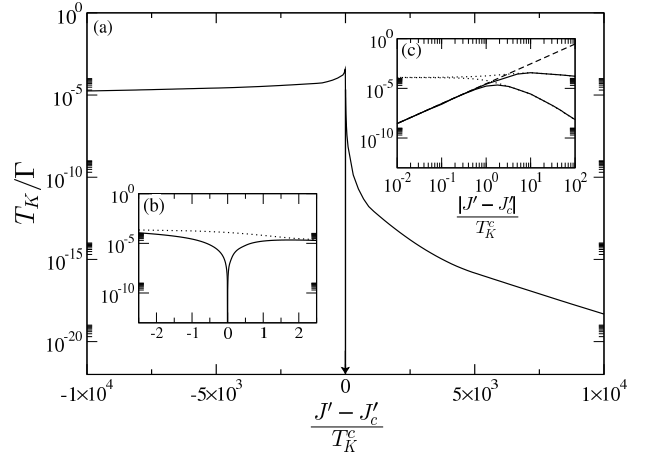}
\caption{\label{Tk trans}
Variation of the Kondo scale $T_{K} \equiv T_{K}^{S}$: $T_K/\Gamma$ \emph{vs} $(J'-J'_{c})/T_{K}^{c}$ (with fixed 
$\tilde{U} =7$ and $\rho J =0.1$, for which $T_{K}^{c}/\Gamma =10^{-4}$).
Inset (b) shows an expanded view in the vicinity of the transition (solid line); inset (c) shows the same results (solid) on a log-log scale, together with the power-law decay $T_{K} \propto |J'_{c}-J'|^{2}$ (dashed line) as the transition is approached. In both insets, $T_{K}^{\chi}$ is also shown (dotted lines), and remains finite at the transition.
}\end{figure}

The obvious first question is how $T_{K}\equiv T_{K}(\tilde{E}_{\Delta})$ vanishes as  $\tilde{E}_{\Delta} =J'_{c}-J' \rightarrow 0$ on approaching the QPT. The answer is
\begin{equation}
\label{scalevan}
T_{K}~\overset{J'\rightarrow J'_{c}\pm}\sim ~ {\cal{A}}|J'_{c}-J'|^{\nu}~~~~~~:\nu =2
\end{equation}
with exponent $\nu =2$, and common amplitudes $\cal{A}$ on approaching $J'_{c}$ from either side.
This is shown in the inset panel (c) to Fig.~\ref{Tk trans} (itself discussed further below). The solid lines therein show 
$T_{K}/\Gamma$ \emph{vs} $|J'_{c}-J'|/T_{K}^{c}$ on a log-log scale, as the transition $J'\rightarrow J'_{c}$ is approached from both sides. The dashed line has a slope $\nu =2$, onto which the $T_{K}$ fall clearly for $|J'_{c}-J'| \lesssim T_{K}^{c}$; so that $T_{K}^{c}$ is as such the `boundary' scale below which Eq.~\ref{scalevan} holds and the critical regime is entered. We add also that an exponent $\nu =2$ reflects the non-trivial nature of the transition: for a first order level-crossing transition, as arises~\cite{3dot1ch:us} if the TQD is instead coupled to a single lead/channel at dot 2 (see Fig.~\ref{dots}), one instead finds~\cite{3dot1ch:us} $\nu =1$. 

  Consider now the QCP itself, $J'=J'_{c}$ (line (f) in Fig.~\ref{td trans}). From the arguments given in Sec.~\ref{PhysPicQPT} we anticipate $S_{\text{imp}}^{QCP}(T)=\ln 2 +S_{\text{imp}}^{1CK}(T)$, where the explicit $\ln 2$ reflects the free pseudospin and $S_{\text{imp}}^{1CK}(T)$ is the entropy for  a \emph{single}-channel 
spin-$\tfrac{1}{2}$ Kondo model. This is indeed  verified in Fig.~\ref{td trans}, where $\ln 2 +S_{\text{imp}}^{1CK}(T)$ is calculated explicitly for a 1-channel Kondo model with $T_{K}^{1CK}$ chosen such that $T_{K}^{1CK} =T_{K}^{c}$. It is seen (dotted line) to coincide perfectly
with $S_{\text{imp}}^{QCP}(T)$ for all $T\lesssim \Gamma$; including the $\ln 4$ entropy plateau reflecting the free pseudopsin \emph{and} spin (the `$SU(2) \times SU(2)$' FP of Sec.~\ref{PhysPicQPT}) , and the crossover at $T \sim T_{K}^{1CK}=T_{K}^{c}$ to the CFP characterised by $S_{\text{imp}}=\ln 2$, where the pseudospin remains free but the spin is wholly quenched.

  Moving away from the QCP, consider now the sequence (b $\rightarrow$ e) in
Fig.~\ref{td trans}, as the transition is progressively approached. 
In all cases $S_{\text{imp}}(T)$ follows the QCP $S_{\text{imp}}^{QCP}(T)$ all the way from
the $\ln 4 $ regime, through the crossover at the common temperature $T \sim T_{K}^{c} =T_{K}^{1CK}$ (which is of course finite at the transition),
to the $\ln 2$ symptomatic of the CFP; before ultimately descending to the stable 2CK FP with
$S_{\text{imp}} =\tfrac{1}{2}\ln 2$ on the scale $T\sim T_{K} \equiv T_{K}^{S}$. Since $T_{K}$ vanishes as the transition is approached, $S_{\text{imp}}(T)$ should exhibit universality in terms of $T/T_{K}$. That it does is shown in the inset to Fig.~\ref{td trans}, where all lines (b-e) collapse to a common scaling curve (solid line).
This scaling is of course characteristic of flow from the vicinity of the \emph{critical} FP to the stable 2CK FP. As such it is thus expected to be distinct from the universal crossover in the \emph{single} spin 2CK model 
(Sec.~\ref{thermo 2CK}), from the $\ln 2$ entropy plateau associated with the free spin-$\tfrac{1}{2}$ \emph{local
moment} FP to the stable 2CK FP; indeed confirmed in the inset to Fig.~\ref{td trans}, where the 
latter is shown as a dashed line. Note also in this regard that lines (a) in Fig.~\ref{td trans}, for
$J'_{c}-J' =\pm T_{K}^{c}$, constitute in effect the boundary between effective single-spin 2CK behavior (arising for $|\tilde{E}_{\Delta}| =|J'_{c}-J'| \gtrsim T_{K}^{c} \equiv T_{K}^{1CK}$) and the critical regime
$|J'_{c}-J'| \lesssim T_{K}^{c}$; as reflected in the direct flow of $S_{\text{imp}}(T)$ from $\ln 4$ to $\tfrac{1}{2}\ln 2$ occurring at that point.

\begin{figure}[t]
\includegraphics*[height=6cm]{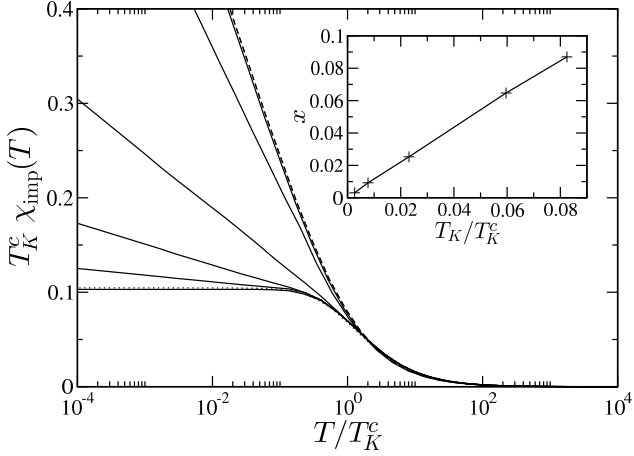}
\caption{\label{chi trans}
Uniform spin susceptibility $T_K^c\chi_{\text{imp}}^{}(T)$ \emph{vs} $T/T_K^c$ for fixed $\tilde{U}=7$ and $\rho J=0.1$, varying $J'$ as: $\tilde{E}_{\Delta}=(J'_{c}-J') =\lambda T_K^{c}$ with $\lambda=10^{n/4}$ and $n=1,0,-1,-2,-3$ (solid lines, in order of decreasing slope). The QCP itself ($\lambda=0$) is the bottom solid line. Also shown (dotted) is a 1CK model chosen such that $T_K^{1CK}=T_K^c$, which describes the QCP; and 
a single-spin 2CK model with $T_K=T_K^c$ (dashed). Near the transition $\chi_{\text{imp}}(T)$ diverges as 
$T_K^c\chi_{\text{imp}}(T)\sim x\ln(T_K/T)$, with amplitude $x\sim T_K/T_K^c$ as shown in the inset
(see also text).
}\end{figure}

 Fig.~\ref{chi trans} shows the $T$-dependence of the uniform spin susceptibility, $T_K^c\chi_{\text{imp}}(T)$ \emph{vs} $T/T_K^c$ (with $T_{K}^{c}/\Gamma = 10^{-4}$ as above), on approaching the transition according to
$\tilde{E}_{\Delta}=J'_{c}-J' =\lambda T_{K}^{c}$; but now with  $\lambda=10^{n/4}$ and $n=1,0,-1,-2,-3$ (solid lines, in order of decreasing slope), showing as such the region where RG flow near the CFP is first observed (roughly between lines (a) and (b) in Fig.~\ref{td trans}). The figure also shows the behavior at the QCP itself ($\lambda =0$, bottom solid line), as well as the spin susceptibility for a single-spin 2CK model (dashed line).

  For the QCP itself, we expect  $\chi_{\text{imp}}^{QCP}(T)= \chi_{\text{imp}}^{1CK}(T)$
from the arguments of Sec.~\ref{PhysPicQPT}; with $\chi_{\text{imp}}^{1CK}(T)$ the spin susceptibility for a single-channel spin-$\tfrac{1}{2}$ Kondo model with Kondo scale $T_{K}^{1CK}\equiv T_{K}^{c}$, such that at $T=0$ in particular $\chi_{\text{imp}}^{1CK}(T=0)$ is characteristically finite. That is indeed verified in Fig.~\ref{chi trans}, where the result for the single-channel Kondo model is shown as a dotted line.

For $n=1$ by contrast, \emph{ie} $J'_{c}-J' \approx 2T_{K}^{c}$, the susceptibility follows essentially perfectly the dashed line; thus being described by a single-spin 2CK model, precisely as arises deep in either 2CK phase of the full TQD model (see Fig.~\ref{chi J'<J}), and characterised by the low-$T$ log divergence
$T_K^c\chi_{\text{imp}}(T) \propto \ln(T_{K}/T)$. On moving closer to the transition (decreasing $n$), the
susceptibilities progressively `fold on'  to the QCP result over an increasingly wider $T/T_{K}^{c}$-range; indeed
even for $J'_{c}-J' =0.1 T_{K}^{c}$ ($n=-4$, not included in Fig.~\ref{chi trans}), the resultant susceptibility is barely distinguishable from the QCP line over most of the $T/T_{K}^{c}$ range shown. As seen from the figure however, in all cases except the QCP itself, the ultimate low-$T$ behavior is the log divergence expected for the stable 2CK FP, $T_{K}^{c}\chi^{}_{\text{imp}}(T) \sim x\ln(T_{K}/T)$; with an amplitude $x$ visible from the gradients in Fig.~\ref{chi trans} and seen to diminish steadily as the transition is approached. While numerical accuracy prevents a definitive determination of $x$ much closer to the transition, extrapolation of the results in Fig.~\ref{chi trans} (as shown in the inset) suggest $x$ vanishes as $x \propto T_{K}/T_{K}^{c}$
($\propto (1-J'/J'_{c})^{2}$ from Eq.~\ref{scalevan}).

  Finally, as discussed only partially above (see Eq.~\ref{scalevan}), we return to the evolution of the Kondo scale shown in Fig.~\ref{Tk trans}; the main panel of which shows $T_{K} \equiv T_{K}^{S}$ \emph{vs} $(J'-J'_{c})/T_{K}^{c}$ as determined from the $T$-dependence of the entropy (Fig.~\ref{td trans}). Inset (b) to Fig.~\ref{Tk trans} shows an expanded view of $T_{K}$ in the vicinity of the transition ($|J'-J'_{c}|/T_{K}^{c} \lesssim 2$), together (dotted line) with the scale $T_{K}^{\chi}$ determined in practice from the spin susceptibility via~\cite{KWW} $T_{K}^{\chi}\chi^{}_{\text{imp}}(T_{K}^{\chi}) = 0.07$ (as employed in Sec.~\ref{thermo 2CK}). While $T_{K} \equiv T_{K}^{S}$ in inset (b) naturally shows the vanishing of the Kondo scale as the QPT is approached, $T_{K}^{\chi}$ is seen to remain finite at the transition, as seen also in Fig.~\ref{Tk scales} of Sec.~\ref{Kondo scales}. This is precisely as it should be. The `true', ultimately vanishing Kondo scale $T_{K}$ is of course evident in the evolution of both $S_{\text{imp}}(T)$ (Fig.~\ref{td trans}) \emph{and} $\chi^{}_{\text{imp}}(T)$ itself (Fig.~\ref{chi trans}). But $T {\times} \chi^{}_{\text{imp}}(T)$ --- from which $T_{K}^{\chi}$ has been defined --- vanishes as $T \rightarrow 0$ (the explicit factor of $T$ `killing' the low-$T$ log divergence in $\chi^{}_{\text{imp}}(T)$ itself). Indeed, as readily inferred from Fig.~\ref{chi trans} (and verified by explicit calculation), $T\chi^{}_{\text{imp}}(T)$ close to the transition ($|J'_{c}-J'|/T_{K}^{c} \lesssim 0.1$ or so) is indistinguishable from the QCP behavior $T\chi_{\text{imp}}^{1CK}(T)$; and as such is naturally characterised by the finite temperature scale $T_{K}^{1CK} \equiv T_{K}^{\chi}$. The above, important distinction between $T_{K}$ and $T_{K}^{\chi}$ is however applicable only close to the transition. As seen clearly from inset (b) to  Fig.~\ref{Tk trans}, the scales $T_{K} \equiv T_{K}^{S}$ and $T_{K}^{\chi}$ coincide away from the transition (in practice for $|J'_{c}-J'|/T_{K}^{c} \gtrsim 2$ or so); as noted and used in Sec.~\ref{thermo 2CK} when considering the model deep in either of the 2CK phases.

\subsection{Dynamics of the transition}\label{dyn trans}
\begin{figure}[t]
\includegraphics*[height=6cm]{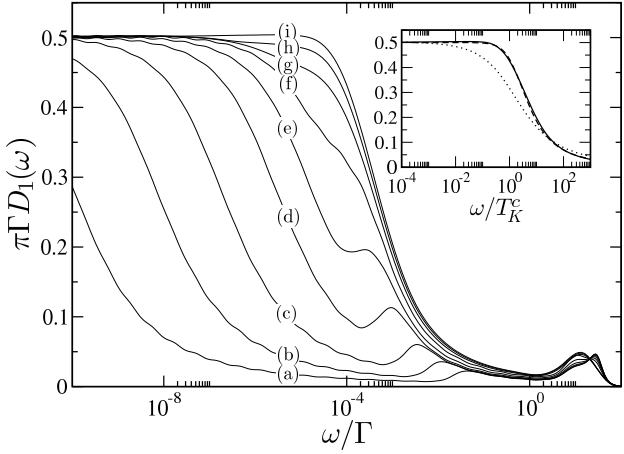}
\caption{\label{Dw trans}
$\pi\Gamma D_1(\omega)$ \emph{vs} $\omega/\Gamma$, varying
$\tilde{E}_{\Delta}= J'_{c}-J'$ as the transition is approached from the $(+)$ parity 2CK phase:  $\tilde{E}_{\Delta}=-\lambda T_K^c$ with $\lambda=10^{n/2}$ and $n=+4\rightarrow -4$ (a$\rightarrow$i respectively). The QCP spectrum ($\lambda =0$) is indistinguishable from case (i).
Inset: QCP scaling spectrum \emph{vs} $\omega/T_{K}^{c}$ (solid line, in practice coinciding with all spectra for $\lambda \lesssim 10^{-2}$). The QCP spectrum is seen to be the scaling spectrum $\tfrac{1}{2}\pi\Gamma D^{1CK}(\omega)$ (dashed line) for a single-channel Anderson impurity model. The dotted line shows the distinct scaling spectrum of a single-spin 2CK model (onto which cases (a)-(c) scale).
}\end{figure}

Here we focus again on the single-particle spectrum $D_{1}(\omega)$ of dot 1 (or equivalently, $D_{3}(\omega)$ for dot 3). The $T=0$ spectrum $\pi\Gamma D_1(\omega)$ \emph{vs} $\omega/\Gamma$ is shown in Fig.~\ref{Dw trans} for $\tilde{U}=7$ and $\rho J=0.1$; varying  $\tilde{E}_{\Delta}= J'_{c}-J'$ on approaching the transition from the $(+)$ parity phase $J' >J'_{c}$, according to $\tilde{E}_{\Delta}=-\lambda T_K^c$ with  $\lambda=10^{n/2}$ and integral $n=+4\rightarrow -4$ for lines a$\rightarrow$i respectively. The spectrum at the transition itself ($\lambda =0$) is indistinguishable from case (i). 

Two particular spectral features should be noted. First, the spectrum at the Fermi level is pinned at $\pi\Gamma D_{1}(\omega =0) =\tfrac{1}{2}$ in all cases, including the transition itself. Second, the width of the Kondo resonance clearly increases as the transition is approached, being of order $T_{K}^{c} \equiv T_{K}^{1CK}$ at the transition (with $T_{K}^{c}/\Gamma =10^{-4}$ as usual); in which sense the vanishing $T_{K}$ scale evident \emph{eg} in the evolution of the entropy (Fig.~\ref{td trans}) does not show up in single-particle dynamics.

That $\pi\Gamma D_{1}(\omega =0)$ should be $\tfrac{1}{2}$ away from the transition is (as in Sec.~\ref{dyn 2CK})
a natural consequence of the stable 2CK FP that ultimately arises (Sec.~\ref{trans thermo}).
To understand why $\pi\Gamma D_{1}(\omega =0)=\tfrac{1}{2}$ at the transition itself, first recall the general result from Sec.~\ref{dyn 2CK} that $\pi\Gamma D_{1}(\omega) =-\pi\rho_{T}^{}\mathrm{Im}t_{L}(\omega)$, with $t_{\alpha}(\omega)$ the t-matrix for scattering of electron states in the $\alpha =L,R$ lead. As discussed in Sec.~\ref{PhysPicQPT}, the effective Hamiltonian for the QCP (Eq.~\ref{Hefftrans2}) is strictly separable into disjoint $\mathcal{T}^{z} =\pm \tfrac{1}{2}$ sectors, the two sectors possessing common eigenvalues. In  this case it is straightforward to show that $t_{\alpha}(\omega)=\tfrac{1}{2}[t_{\alpha}(\mathcal{T}^{z}=+\tfrac{1}{2}; \omega)+t_{\alpha}(\mathcal{T}^{z}=-\tfrac{1}{2}; \omega)]$, where $t_{\alpha}(\mathcal{T}^{z};\omega)$ is the $t_{\alpha}$-matrix calculated for a \emph{fixed} $\mathcal{T}^{z}$ in the QCP Hamiltonian Eq.~\ref{Hefftrans2} (with $t_{L}(\mathcal{T}^{z}=\pm \tfrac{1}{2};\omega) = t_{R}(\mathcal{T}^{z}=\mp \tfrac{1}{2};\omega)$ such that $t_{L}(\omega) =t_{R}(\omega)$ overall, as symmetry requires generally). Consider then $t_{L}(\omega)$.
As explained in Sec.~\ref{PhysPicQPT}, at the critical fixed point the spin-$\tfrac{1}{2}$ $\hat{\textbf{S}}$ is quenched in a \emph{one}-channel fashion, by AF coupling to the $L$-lead for $\mathcal{T}^{z}=+\tfrac{1}{2}$, but to the $R$-lead for $\mathcal{T}^{z}=-\tfrac{1}{2}$. In consequence, $t_{L}(\mathcal{T}^{z}=+\tfrac{1}{2}; \omega =0)$ is equivalently the t-matrix, $t^{1CK}(\omega =0)$, for a \emph{one}-channel AF spin-$\tfrac{1}{2}$ Kondo model. But $t_{L}(\mathcal{T}^{z}=-\tfrac{1}{2}; \omega =0)=0$ by contrast, since for $\mathcal{T}^{z}=-\tfrac{1}{2}$ the spin is quenched by coupling to the $R$-lead, while the $L$-lead is entirely decoupled from it (so no spin-scattering of electrons in the $L$-lead can occur). Hence, 
$t_{L}(\omega =0) \equiv \tfrac{1}{2}t^{1CK}(\omega =0)$. For a one-channel spin-$\tfrac{1}{2}$ Kondo model itself,
$-\pi\rho_{T}^{}\mathrm{Im}t^{1CK}(\omega) \equiv \pi\Gamma D^{1CK}(\omega)$, with $D^{1CK}(\omega)$ the single-particle spectrum of a one-channel, single-level Anderson impurity model in the singly-occupied Kondo regime of the model (as is physically obvious, but follows more formally via arguments directly analogous to those given in Sec.~\ref{dyn 2CK}). Overall we thus infer that $\pi\Gamma D_{1}(0) = -\pi\rho^{}_{T}\mathrm{Im}t_{L}(0) \equiv \tfrac{1}{2}\pi\Gamma D^{1CK}(0)$. But $\pi\Gamma D^{1CK}(0) =1$ by virtue of the Friedel sum rule~\cite{hewson}; whence $\pi\Gamma D_{1}(0)=\tfrac{1}{2}$ arises, as indeed found, see Fig.~\ref{Dw trans}.

While the arguments above apply to $\omega =0$ (and as such the CFP), one naturally 
expects the frequency dependence of the Kondo resonance in $\pi\Gamma D_{1}(\omega)$ at the QCP to be that of $\tfrac{1}{2}\pi\Gamma D^{1CK}(\omega)$. That this is so is demonstrated in the inset to Fig.~\ref{Dw trans}. The scaling spectrum $\tfrac{1}{2}\pi\Gamma D^{1CK}(\omega)$ for the Anderson model (as a function of $\omega/T_{K}^{1CK}$) is shown as a dashed line, and compared to $\pi\Gamma D_{1}(\omega)$ \emph{vs} $\omega/T_{K}^{c}$ for the full TQD model at the QCP (solid line); the two coincide perfectly. In particular, the low-frequency spectral behavior ($|\omega|/T_K^c\ll 1$) at the transition is thus of quadratic Fermi liquid form~\cite{hewson}, 
\begin{equation}
\label{QCPspec}
\pi\Gamma D_1(\omega)~\sim ~ \tfrac{1}{2}\left[1-a'(\omega/T_K^c)^{2}\right]
\end{equation}
with $a'$ a constant of order unity. This is of course in marked contrast to the behavior arising deep in either of the 2CK phases (Sec.~\ref{dynamicsresults}). The scaling spectrum in that case is also shown in Fig.~\ref{Dw trans}
(inset, dotted line), and instead exhibits the characteristic square-root frequency dependence of Eq.~\ref{lowfreqD}.

Cases (a)-(c) in Fig.~\ref{Dw trans} are relatively deep in the ($+$)-parity 2CK phase ($|\tilde{E}_{\Delta}| \geq 10T_{K}^{c}$). Their dynamics are thus in essence those of the single-spin 2CK model discussed in Sec.~\ref{dyn 2CK}: when scaled in terms of their Kondo scale $T_{K}$, the spectra all `collapse' onto the 2CK scaling spectrum shown in the inset (dotted), departure from such occurring only by non-universal scales $\omega \sim |\tilde{E}_{\Delta}| $ (where a subsidiary peak arises, see main figure). Close to the transition by contrast
--- exemplified by case (i) in Fig.~\ref{Dw trans} ($|\tilde{E}_{\Delta}| =10^{-2} T_{K}^{c}$) --- the single-particle dynamics are indistinguishable from that of the QCP, as is physically natural. Equally naturally, spectra (d)-(h) 
do not conform to either limiting form (QCP or 2CK), but instead represent crossover behavior between the two.

\begin{figure}[t]
\includegraphics*[height=6cm]{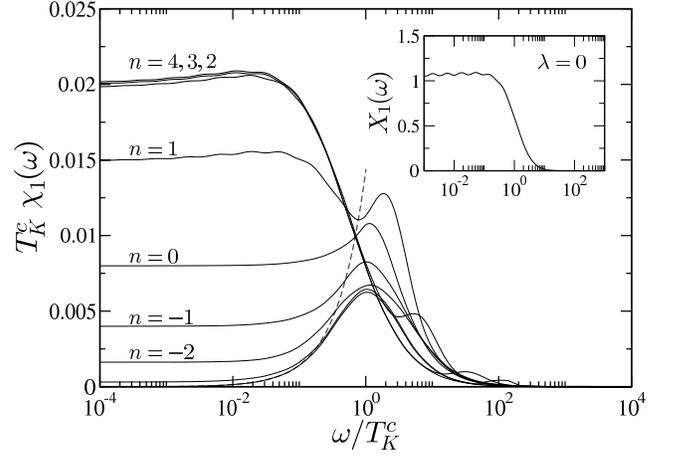}
\caption{\label{X1 trans}
Local dynamic susceptibility $T_K^c\chi_1^{}(\omega)$ \emph{vs} $\omega/T_K^c$, varying $\tilde{E}_{\Delta}=J'_{c}-J'$ on approaching the transition from the $(-)$-parity 2CK phase:
$\tilde{E}_{\Delta}=\lambda T_K^c$ with $\lambda=10^{n/2}$ and $n=+4\rightarrow -4$.
$T_{K}^{c}\chi_1^{}(0)$ vanishes as the transition is approached.
The dashed line, $T_K^c \chi_1^{}(\omega)\propto \omega$, shows the low-$\omega$ Fermi liquid behavior of the single-channel Anderson model arising at the QCP ($\lambda=0$). Inset: $X_1(\omega)=\chi_1^{}(\omega)/(2\omega [\text{Re}\langle\langle S_1^z;S_1^z \rangle \rangle_{\omega=0}]^{2})$ \emph{vs} $\omega/T_K^c$ at the transition ($\lambda=0$), showing recovery of the Korringa-Shiba relation $X_1(\omega = 0)=1$.
}\end{figure}

Finally, we consider the $T=0$ local dynamic susceptibility for dot 1 in the vicinity of the transition. Fig.~\ref{X1 trans} shows $T_K^c\chi_1^{\phantom{\dagger}}(\omega)$ \emph{vs} $\omega/T_K^c$, varying $\tilde{E}_{\Delta} =J'_{c}-J'$ on approaching the transition from the ($-$)-parity 2CK phase $J'<J'_{c}$, according to $\tilde{E}_{\Delta}=\lambda T_K^c$ with $\lambda=10^{n/2}$ and $n=+4\rightarrow -4$.
The cases $n=2,3,4$ (with $\tilde{E}_{\Delta} \geq 10T_{K}^{c}$) are quite deep in the 2CK phase (where the $T_{K}$ scale is roughly constant, see Fig.~\ref{Tk trans}); so they are essentially equivalent to the 2CK scaling curves of Fig.~\ref{dyn susc} (inset).

As the transition is approached however, the height of the $\omega \rightarrow 0$ plateau in Fig.~\ref{X1 trans} is seen steadily to diminish; the behavior found being of form $T_K^c\chi_1^{\phantom{\dagger}}(\omega=0)\sim T_K/T_K^c$, with the Kondo scale $T_{K} \rightarrow 0$. This is readily understood: as mentioned in Sec.~\ref{dynamicsresults}
it is known~\cite{2ck:dyn_anders,2ck:dyn_bradley} that at the 2CK FP, $\chi_{1}^{}(0)$ plateaus at a constant, itself proportional to the slope of the low-temperature log divergence  of $\chi^{}_{\text{imp}}(T)$;
and in Fig.~\ref{chi trans} we showed the latter to vanish  $\propto T_{K}\rightarrow 0$.
Hence  $\chi_1^{\phantom{\dagger}}(0) =0$ at the transition itself. As expected from the nature of the QCP, the leading low-$\omega$ dependence of $\chi_1^{\phantom{\dagger}}(\omega)$ is then the Fermi liquid behavior characteristic of the single-channel Anderson model\cite{hewson,shiba}, $\chi_1^{\phantom{\dagger}}(\omega) \propto \omega$, shown in Fig.~\ref{X1 trans}\cite{note:lowomega} (dashed line); and $\chi_1^{\phantom{\dagger}}(\omega)$ is also seen to contain an absorption centered on $\omega =T_{K}^{c}$ ($\equiv T_{K}^{1CK}$), likewise characteristic of the Anderson model\cite{hewson,shiba}.

The inset to Fig.~\ref{X1 trans} shows $X_1(\omega)=\chi_1(\omega)/(2\omega [\text{Re}\langle\langle S_1^z;S_1^z \rangle \rangle_{\omega=0}]^{2})$ \emph{vs} $\omega/T_K^c$ at the transition itself. For the single-channel Anderson model, the $\omega \rightarrow 0$ behavior of $X_1(\omega)$ is given exactly by the Korringa-Shiba relation\cite{hewson,shiba} $X_1(\omega = 0)=1$; seen from the figure to be well satisfied in practice (to within a few $\%$), again confirming the physical picture of the CFP discussed in Sec.~\ref{PhysPicQPT}.


\section{Reduced model for the transition}\label{red}
In the preceding sections, the effective low-energy model Eq.~\ref{Heff trans} (or Eq.~\ref{Hefftrans1}) has been 
important in understanding the behavior of the full TQD system in the vicinity of the QPT. We have also performed direct NRG calculations on the effective low-energy model itself, varying independently the bare parameters $J_{A}, J_{B}, J_{\text{mix}}$ and $E_{\Delta}$ (or $\tilde{E}_{\Delta}$); and now comment briefly on the results of such.
Specifically, we here consider explicitly the effective model with $J_{A}=0=J_{B}$, which we have confirmed describes the same physics as arises with generically non-zero $J_{A}$ or $J_{B}$  (reflecting the fact that the key element in the effective Hamiltonian Eq.~\ref{Heff trans} is the pseudospin raising/lowering term,
$H_{\text{mix}}$ below, which in switching the parity of the TQD states effectively drives the QPT).

\begin{figure}[t]
\includegraphics*[height=6cm]{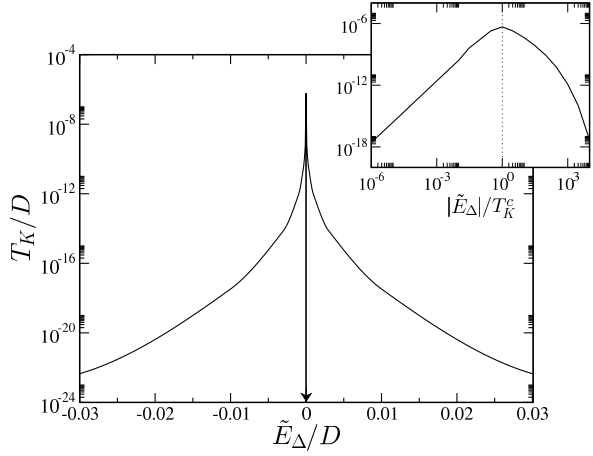}
\caption{\label{Tk red}
Evolution of the 2CK scale for the reduced model Eq.~\ref{red H}: $T_K/D$ \emph{vs} $\tilde{E}_{\Delta}/D$ (with $D$ the conduction bandwidth), for fixed $\rho J_{\text{mix}}=0.075$. Inset: $T_K/D$ \emph{vs} $|\tilde{E}_{\Delta}|/T_K^c$ on logarithmic axes, with $T_K\sim |\tilde{E}_{\Delta}|^2$ on approaching the transition.
}
\end{figure}
 
We thus consider the reduced model $H_{\text{red}}=H_0+H_{\text{mix}}+H_{\text{mag}}$, with $H_{0}$

the Hamiltonian for the leads and
\begin{subequations}\label{red H}
\begin{align}
H_{\text{mix}}&=J_{\text{mix}}(\hat{\mathcal{T}}^{+}+\hat{\mathcal{T}}^{-})\hat{\textbf{S}}\cdot(\hat{\textbf{s}}_{L0} - \hat{\textbf{s}}_{R0}),
\label{mix}\\
\label{mag}
H_{\text{mag}}&=\tilde{E}_{\Delta}\hat{\mathcal{T}}^z.
\end{align}
\end{subequations}
The underlying FPs of $H_{\text{red}}$ are readily inferred. The $SU(2)\times SU(2)$ FP corresponds to $J_{\text{mix}}=0=\tilde{E}_{\Delta}$ in Eq.~\ref{red H}, and as such to both a free spin and a free pseudospin; this of course is the high-temperature FP of $H_{\text{red}}$, generating \emph{eg} the associated $\ln 4$ entropy. Two LM FPs also arise, corresponding formally to $J_{\text{mix}}=0$ in Eq.~\ref{red H} (and hence a free spin), and $\tilde{E}_{\Delta} \rightarrow \pm\infty$. For $\tilde{E}_{\Delta} \rightarrow +\infty$, the $\mathcal{T}^{z} =-\tfrac{1}{2}$ pseudospin component (\emph{ie} the ($-$)-parity TQD doublet) is frozen out, and we designate this as the `LM($-$)' FP; while for $\tilde{E}_{\Delta} \rightarrow -\infty$ the $\mathcal{T}^{z} =+\tfrac{1}{2}$ component is frozen out, corresponding to a LM($+$) FP.

\begin{figure}[t]
\includegraphics*[height=6cm]{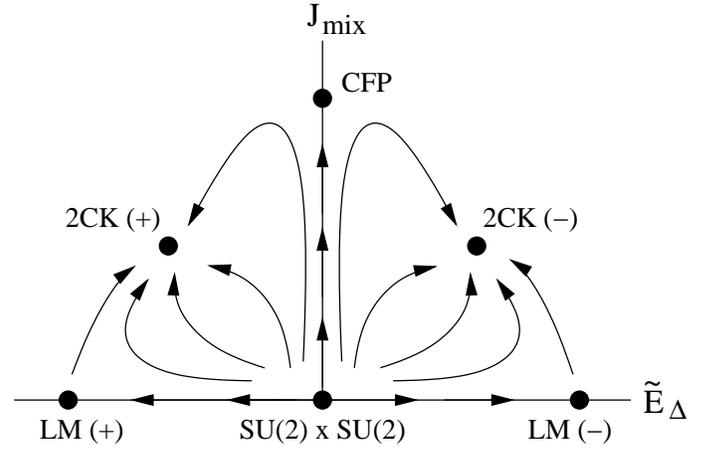}
\caption{\label{RG flow}
Schematic illustration of RG flow between the FPs (circles) of the reduced model. See text for discussion.
}\end{figure}

The critical FP arises for $\tilde{E}_{\Delta}=0$. In this case (see Sec.~\ref{PhysPicQPT}), a trivial rotation of the pseudospin axes gives the QCP Hamiltonian
$H_{\text{red}}^{\text{QCP}}=H_{0}+2J_{\text{mix}}\hat{\mathcal{T}}^{z}\hat{\textbf{S}}\cdot(\hat{\textbf{s}}_{L0} - \hat{\textbf{s}}_{R0})$, with the quantum number $\mathcal{T}^{z}$ conserved and the Hilbert space thus separable into disjoint $\mathcal{T}^{z} =\pm \tfrac{1}{2}$ sectors. As discussed in Sec.~\ref{PhysPicQPT}, the CFP itself corresponds to the spin being quenched in a one-channel fashion by coupling to the $L$ lead in the $\mathcal{T}^{z} = +\tfrac{1}{2}$ sector, and to the $R$ lead for $\mathcal{T}^{z} = -\tfrac{1}{2}$. More formally, on writing
$H_{\text{red}}^{\text{QCP}}=H_{0}+2\hat{\mathcal{T}}^{z}\hat{\textbf{S}}\cdot(J_{\text{mix}}^{L}\hat{\textbf{s}}_{L0} - J_{\text{mix}}^{R}\hat{\textbf{s}}_{R0})$  --- with $J_{\text{mix}}^{L} = J_{\text{mix}}^{R} = J_{\text{mix}}>0$ in the `bare' $H_{\text{red}}^{\text{QCP}}$ --- the CFP corresponds to $J_{\text{mix}}^{L} \rightarrow \infty$ and $J_{\text{mix}}^{R}=0$ in the $\mathcal{T}^{z}=+\tfrac{1}{2}$ sector, with
$J_{\text{mix}}^{L} =0$ and $J_{\text{mix}}^{R}\rightarrow \infty$ for $\mathcal{T}^{z}=-\tfrac{1}{2}$.

 Finally, for $\tilde{E}_{\Delta} \neq 0 \neq J_{\text{mix}}$, the higher lying component of the pseudospin is naturally frozen out on sufficiently low energy scales $T \ll |\tilde{E}_{\Delta}|$. Virtual excitations to the upper pseudospin sector still arise, however; and treating the mixing term Eq.~\ref{mix} perturbatively  (strictly valid for large $|\tilde{E}_{\Delta}|$) gives $H_{\text{red}}^{\text{eff}}=H_0+H_{\text{hyb}}^{\text{eff}}$, with 
\begin{equation}
\label{red 2ck}
H_{\text{hyb}}^{\text{eff}}=\left ( \frac{J_{\text{mix}}^{2}} {2E_{\Delta}}\right ) \hat{\textbf{S}}\cdot(\hat{\textbf{s}}_{L0} + \hat{\textbf{s}}_{R0}).
\end{equation}
This is a model of 2CK form in the reduced Hilbert space of the lowest pseudospin component (($+$)-parity for 
$\tilde{E}_{\Delta} <0$ and ($-$)-parity for $\tilde{E}_{\Delta} >0$), for which the ultimate stable FP is of course the infrared 2CK FP.

The overall low-energy FP structure of the reduced model is thus indeed that of the full TQD model (we discuss its RG flow diagram below). On studying Eq.~\ref{red H} directly with NRG the transition occurs as expected at $\tilde{E}_{\Delta}=0$, and the above FPs are indeed observed \emph{eg} in the $T$-dependence of thermodynamics.  

Fig.~\ref{Tk red} shows the resultant 2CK scale $T_K/D$ \emph{vs} $\tilde{E}_{\Delta}/D$ for systems with common $\rho J_{\text{mix}}=0.075$. $T_K$ is determined in the usual way from the entropy, and found to be non-zero for any $\tilde{E}_{\Delta}\ne 0$ (Fig.~\ref{Tk red}); and we identify $T_K^c$ ($\propto \exp(-1/\rho J_{\text{mix}})$) as the crossover scale from the $SU(2)\times SU(2)$ FP to the CFP. The inset to Fig.~\ref{Tk red}  shows $T_K/D$ \emph{vs} $|\tilde{E}_{\Delta}|/T_K^c$ on logarithmic axes. Just like Fig.~\ref{Tk trans} for the full TQD model, the point $\tilde{E}_{\Delta}=T_K^c$ can be seen to separate two distinct regimes. For $|\tilde{E}_{\Delta}|\gg T_K^c$, systems may be considered `deep' in the 2CK phase (which situation constitutes the majority of the main panel of Fig.~\ref{Tk red}, and indeed the majority of the parameter space of Eq.~\ref{red H}). For $\tilde{E}_{\Delta}\ll T_K^c$ by contrast, systems flow close to the CFP, and are `near' the transition. The vanishing $T_K$ scale associated with the transition is found to be characterized by a power-law decay $T_K\sim |\tilde{E}_{\Delta}|^{\nu}$ with $\nu=2$, as found in the full TQD model (Sec.~\ref{trans thermo}). We also add that since the reduced model Eq.~\ref{red H} lacks a direct coupling of form $\hat{\textbf{S}}\cdot(\hat{\textbf{s}}_{L0} + \hat{\textbf{s}}_{R0})$ correlated to the pseudospin (\emph{cf} Eq.~\ref{Heff trans}),
the variation of $T_K$ is symmetric in $\tilde{E}_{\Delta}$; as evident in Fig.~\ref{Tk red}.

A schematic illustration of the RG flow is given in Fig.~\ref{RG flow}, in the ($\tilde{E}_{\Delta}, J_{\text{mix}}$)-plane. The FPs are indicated by circles, the arrows representing RG flow between them. At the highest temperatures, the system is always described by a free spin and a free pseudospin (the $SU(2)\times SU(2)$ FP, $J_{\text{mix}}=\tilde{E}_{\Delta}=0$ in Eq.~\ref{red H}). 

Considering now the case of $J_{\text{mix}}=0$, the spin always remains free; but the upper pseudospin component 
is frozen out for any $\tilde{E}_{\Delta}\ne 0$, whence as $T\rightarrow 0$ the system is described by either a $(+)$ or $(-)$ parity LM FP. By contrast, for $\tilde{E}_{\Delta}=0$, any $J_{\text{mix}}\ne 0$ drives the system to the CFP as $T\rightarrow 0$. In the general case where $|\tilde{E}_{\Delta}|\ne 0$ and $J_{\text{mix}}\ne 0$, the system always flows ultimately to a 2CK FP (the parity of which depends on which component of the pseudospin lies lowest, \emph{ie} whether $\tilde{E}_{\Delta}>0$ or $<0$). In this case RG flow first approaches the LM FPs if $|\tilde{E}_{\Delta}|\gg T_K^c$, or the CFP for $|\tilde{E}_{\Delta}|\ll T_K^c$, before finally flowing to the appropriate stable 2CK FP.

  And as evident from the results of Sec.~\ref{QPT} (\emph{eg} Fig.~\ref{td trans}), the RG flows depicted schematically in Fig.~\ref{RG flow} are just those arising at low-energies in the full TQD model, in the vicinity of, and at, the quantum phase transition.


\section{Conclusions}\label{concs}
We have studied in this paper what is arguably the simplest 2-channel model in which to understand 
the consequences of local frustration arising from internal degrees of freedom: a 
triple quantum dot ring, with dots mutually coupled by antiferromagnetic exchange interactions, and tunnel-coupled symmetrically to two metallic leads. While important aspects of the general model have been considered 
before\cite{zitkoTQD2ch,zitkoFL_NFL}, the underlying physics is found to be 
even richer than hitherto uncovered. 

Two distinct 2CK phases arise due to the mirror symmetry in the problem, each displaying classic non-Fermi liquid properties of the 2CK fixed point below a characteristic 2-channel Kondo scale. But although 2-channel Kondo physics predominates in the underlying parameter space, the parity-distinct nature of the  2CK phases means that a quantum phase transition between them occurs. Driven by varying the interdot exchange couplings, occurring at the point of inherent magnetic frustration, and characterised by a nontrivial quantum critical point which we have explicitly  identified and analysed, the transition provides a striking example of the subtle interplay between internal spin and orbital degrees of freedom.


\begin{acknowledgments}
Helpful discussions with M. Galpin, C. Wright and T. Jarrold are gratefully acknowledged. We thank EPSRC (UK) for financial support, under grant EP/D050952/1.
\end{acknowledgments}


\appendix*
\section{}
Here we specify the parity operator $\hat{P}_{i,j}$ which exchanges the labels $i$ and $j$ for an arbitrary pair of orbitals $i,j$. First we transform canonically to even ($e$) and odd ($o$) orbitals,
\begin{subequations}\label{even/odd}
\begin{align}
c^{\dagger}_{e\sigma}=&\tfrac{1}{\sqrt{2}}(c^{\dagger}_{i\sigma}+c^{\dagger}_{j\sigma})\\
c^{\dagger}_{o\sigma}=&\tfrac{1}{\sqrt{2}}(c^{\dagger}_{i\sigma}-c^{\dagger}_{j\sigma}).
\end{align}
\end{subequations}
Exchanging $i \leftrightarrow j$ clearly has no effect on $c^{\dagger}_{e\sigma}$, but $
c^{\dagger}_{o\sigma} \leftrightarrow -c^{\dagger}_{o\sigma}$. 
With $\hat{n}_o=\hat{n}_{o\uparrow}+\hat{n}_{o\downarrow}$ the odd-orbital number operator we thus define:
\begin{equation}\label{P_o}
\hat{P}_{i,j}=2(\hat{n}_o-1)^2-1
\end{equation}
$\hat{P}_{i,j}$ is self-adjoint and involutory ($\hat{P}_{i,j}^{2}=1$), and hence has eigenvalues $\pm 1$ only
(specifically $P_{i,j}=+1$ for $n_o=0,2$ and $P_{i,j}=-1$ for $n_o=1$). From Eqs.~\ref{even/odd} and~\ref{P_o}, it follows that 
\begin{subequations}\label{P comm}
\begin{align}
[\hat{P}_{i,j}&,c^{\dagger}_{e\sigma}]=0,\\
\{\hat{P}_{i,j}&,c^{\dagger}_{o\sigma}\}=0~,
\end{align}
\end{subequations}
and hence from Eqs.~\ref{even/odd} and~\ref{P comm}
\begin{subequations}\label{permute ij}
\begin{align}
\hat{P}_{i,j}c^{\dagger}_{i\sigma}&=c^{\dagger}_{j\sigma}\hat{P}_{i,j}\\
\hat{P}_{i,j}c^{\dagger}_{j\sigma}&=c^{\dagger}_{i\sigma}\hat{P}_{i,j}~,
\end{align}
\end{subequations}
showing that $\hat{P}_{i,j}$ indeed permutes the $i,j$ labels. Using Eq.~\ref{even/odd} in Eq.~\ref{P_o} it is straightforward (if lengthy) to show that
\begin{equation}
\label{parityop_app}
\begin{split}
\hat{P}_{i,j}~=~&2(\hat{\textbf{I}}_i\cdot\hat{\textbf{I}}_j -\hat{\textbf{S}}_i\cdot\hat{\textbf{S}}_j)~ + 
\sum_{i=\{i,j\},\sigma}\tfrac{1}{2}\hat{n}_{i\sigma}(\hat{n}_{i-\sigma}-1)\\
&+ \sum_{\sigma}(c_{i \sigma}^{\dagger} c_{j \sigma}^{\phantom{\dagger}}+\text{H.c.})(2-\hat{n}_i-\hat{n}_j)+\tfrac{1}{2}
\end{split}
\end{equation} 
where $\hat{\textbf{I}}_i$ is an isospin operator (with components  $\hat{I}_{i}^{z}=\tfrac{1}{2}(\hat{n}_{i}-1)$ and $\hat{I}_{i}^{+}=c_{i\uparrow}^{\dagger}c_{i\downarrow}^{\dagger}$
and $\hat{I}_{i}^{-}=(\hat{I}_{i}^{+})^{\dagger}$).

If a Hamiltonian is invariant to the permutation $i \leftrightarrow j$, then $[H,\hat{P}_{i,j}]=0$,
parity is conserved, and all states can thus be classified according to it. The isolated trimer Hamiltonian
$H_{\text{tri}}$ (Eq.~\ref{Hfull}) is clearly invariant under interchange of the $1 \leftrightarrow 3$ dot levels,
$\hat{P}_{1,3}$. The full lead-coupled Hamiltonian is by contrast invariant to simultaneous interchange of 1 and 3 labels \emph{and} left and right leads, the larger (`overall $L\leftrightarrow R$') symmetry of $H$ implying that the relevant involutory permutation operator is 
\begin{equation}\label{P_LR}
\hat{P}_{L,R}=\hat{P}_{1,3}\prod_{\textbf{k}}\hat{P}_{L\textbf{k},R\textbf{k}}~,
\end{equation} 
such that $[H,\hat{P}_{L,R}]=0$.



\begin{thebibliography}{93}
\bibitem[{not({\natexlab{a}})}]{note:jsimu}
\bibinfo{note}{The essential physics is of course the same whether $J\sim U$ or
  $\ll U$, since the dots remain singly-occupied; and we consider here a
  relatively large $J/U \sim {\cal{O}}(1)$ largely for numerical convenience,
  albeit noting that quantum dots in graphene typically
  have~\cite{graph:baskaran} $J/U \sim \tfrac{1}{3}$.}
\bibitem[{not({\natexlab{b}})}]{note:ln8}
\bibinfo{note}{In practice the systems in Fig.~\ref{thermo J'<J} flow directly
  to the LM FP from the FO FP, with no $\ln(8)$ entropy plateau since there is
  not a wide scale separation between $U$ and $J$.}
\bibitem[{not({\natexlab{c}})}]{note:Gs}
\bibinfo{note}{We have developed a scheme for an accurate determination of
  $G_{s}(\omega)$ itself, in the spirit of [\onlinecite{UFG}] and analogous to
  the determination of $G_{1}(\omega)$ via
  Eqs.~(\ref{self-energy},\ref{dyson}).}
\bibitem[{not({\natexlab{d}})}]{note:lowomega}
\bibinfo{note}{For $\tilde{E}_{\Delta} \lesssim T_{K}^{c}$, the leading
  low-$\omega$ behavior of $T_K^c\chi_1^{\phantom{\dagger}}(\omega)$ is in fact
  well described by $T_K^c\chi_1^{\phantom{\dagger}}(\omega)\sim
  T_K^c\chi_1^{\phantom{\dagger}}(0)+\alpha\omega$ (with $\alpha$ a constant).}
\end{thebibliography}
\end{document}